# Plausible Is Not Enough: Trust and Evidence in AI-Authored Pull Request Review

Md Shamimur Rahman, Khairul Alam, Banani Roy, Chanchal K. Roy

Department of Computer Science, University of Saskatchewan, Saskatoon, Saskatchewan, Canada.

*Corresponding author(s). E-mail(s): mdr614@usask.ca;
Contributing authors: kha060@usask.ca; banani.roy@usask.ca; croy@cs.usask.ca;

**Abstract**

Code review is the checkpoint where AI-authored pull requests (AIPRs) must earn reviewer trust before integration, yet it remains unclear whether reviewers apply the same decision logic they use for human-authored pull requests (HPRs). We report a mixed-method questionnaire survey of 239 practitioners from 31 countries in which respondents answered matched questions about HPRs and AIPRs, enabling within-respondent comparisons of trust, evidence requirements, and verification priorities. Respondents reported lower initial trust in AIPRs, with 64.9% rating them lower than HPRs, and placed less emphasis on author-centered cues such as track record and subsystem familiarity, highlighting instead human prompter expertise and bounded change size. At approval, respondents more often identified their own reasoning through critical cases as sufficient evidence for AIPRs (55.2%) than HPRs (28.5%), while relying less on test coverage as sufficient evidence. Overall, 61.1% reported higher evidence requirements for AIPRs, and blocking approval under residual uncertainty rose from 19.4% for HPRs to 43.9% for AIPRs. Respondents also gave greater priority to test inspection, context checking, and local execution, and more often selected partial reimplementation when verifying non-trivial logic. Unclear ownership became a more prominent reason to pause AIPR review (29.3% versus 14.6% for HPRs), while open-ended responses highlighted the difficulty of validating plausible code without reliable project context and human accountability. These findings show that AIPR review is not only about generated-code correctness. It is an evidence-ownership problem: reviewers need a defensible chain from generated change to validated behavior and maintainable ownership. Review tools and agentic workflows should therefore surface provenance, distinguish generated claims from human-validated evidence, and make unresolved uncertainty visible, helping reviewers decide when plausible code is trustworthy enough to merge.

## 1 Introduction

Peer code review is central to quality assurance and collaboration in modern software development. Before integration, reviewers must decide whether a submitted change is understandable, justified, and safe enough to merge. Prior work shows that code review helps detect defects, technical debt, and code smells [1–5], while supporting knowledge sharing, project conventions, and developer coordination [6–9]. In pull-based development, these judgments are mediated through Pull Requests (PRs), where reviewers must understand rationale [10, 11], assess quality and integration risk [5, 8], judge test adequacy [9, 12], and provide useful feedback [8, 13]. Thus, even for Human-authored PRs (HPRs), review is not mechanical inspection, but a judgment process under incomplete information, limited time, and varying trust.

AI-based code generation is changing the conditions under which review judgments are made. Earlier tools such as GitHub Copilot [14] mainly assisted developers during coding, while newer coding agents, including

Copilot Cloud [15], Claude Code [16], and Codex Web [17], can inspect repositories, modify code, create branches, and open or update PRs with varying human direction. Studies of AI-assisted programming show that developers use code-generating models to accelerate implementation and explore alternatives, but still must inspect, debug, and adapt generated code before relying on it [18–20]. Recent work on agentic PRs shows that AI-generated or AI-assisted PRs are already appearing in real repositories, with diverse change types, outcomes, and failure modes [21–24]. These developments make code review a critical point of human oversight in AI-assisted software development.

A central question is whether reviewers evaluate AI-authored PRs (AIPRs) using the same trust model they apply to HPRs. In conventional PR review, a submitted change is usually tied to an identifiable human contributor. Reviewers can use social and project-history signals, including prior contribution record [25–27], subsystem familiarity [28], and responsiveness during review [29], to calibrate trust and judge when evidence is sufficient. When a PR is perceived as AI-authored or agent-generated, these signals become less direct. The code may appear plausible and pass tests, yet reviewers may remain uncertain whether the submitter understands the implementation, whether the rationale is reliable, whether relevant edge cases were considered, and whether generated tests or explanations provide meaningful assurance. Prior work on AI-generated code raises related concerns about correctness, robustness, and security [18, 30–32]. From a trust-in-automation perspective, AIPRs therefore create a calibration problem: reviewers must decide when to rely on AI-produced artifacts, when to demand additional evidence, and when to block approval because trust is not yet warranted [33–35].

Although Modern Code Review (MCR) and AI-assisted programming have both been extensively studied, we still know little about how *perceived authorship* changes reviewers' approval reasoning. Code review research has examined review quality, feedback usefulness, reviewer expertise, participation, PR acceptance, and confusion [5, 8, 10, 13, 36, 37]. Additionally, AI coding research has examined how developers use code-generation tools, validate and repair generated suggestions, and how agentic PRs appear in repository traces [18–20, 22, 24, 38]. Yet these have not explained whether reviewers change their initial trust, trust cues, evidence thresholds, verification strategies, and tolerance for unresolved uncertainty when the same review decision concerns a human-authored rather than an AI-authored contribution. This gap is important for the safety and scalability of AI-assisted development, as human reviewers must be able to assess AI-generated code accurately without placing excessive trust in seemingly correct outputs or unnecessarily rejecting useful contributions.

In this paper, we study AIPR review as a problem of *trust calibration and evidence requirements*. We define *perceived authorship* as the reviewer's interpretation of a PR as primarily human-authored or AI-authored at review time, independent of exact line-level provenance. This framing reflects mixed-authorship practice, where humans may prompt, revise, submit, or validate substantially AI-generated code. We conducted a mixed-method questionnaire survey of 239 practitioners with experience in pull-based development, code review, and AI-authored or agent-generated PRs. Each respondent answered matched questions about HPRs and AIPRs, so differences in trust formation, evidence requirements, uncertainty handling, verification depth, and effort allocation can be examined within the same person. Respondents reported that perceived AI authorship changes the basis on which they decide whether a PR is safe to approve. They sought stronger connections between AI-generated code, human verification, and long-term maintenance responsibilities, recognizing that passing tests, plausible explanations, and familiar review practices alone were insufficient. This shift was evident in trust formation, evidence thresholds, tolerance for uncertainty, and verification effort. However, the differences were selective, as respondents changed how they judged assurance and accountability while keeping the technical checks they apply to human-authored changes. Taken together, the responses suggest that AIPR review requires reviewers to defend the approval decision more explicitly than in conventional HPR review. The central challenge, therefore, is not only assessing whether generated code works, but determining whether the available evidence makes the change trustworthy enough to merge. Overall, our contributions are as follows:

- **Empirical evidence on AIPR review reasoning.** We provide a within-respondent comparison of how practitioners evaluate HPRs and AIPRs across the full approval decision, from initial trust and trust cues to evidence requirements, handling of unresolved uncertainty, and reported verification practices.
- **Conceptual framing of AIPR review.** We interpret the findings as an evidence-ownership problem, in which reviewers need to know who can substantiate the claims behind a generated change, and describe understanding and ownership debt as a possible long-term maintenance risk.
- **Implications for review practice and tooling.** We derive practical implications for risk-sensitive AIPR review, emphasizing independent validation, code provenance, and human ownership, and propose evaluating coding agents based on the verification burden they impose on reviewers.

# 2 Background and Related Work

## 2.1 Review Judgment and Authorship Cues

MCR involves assessing the evidence supporting a code change beyond verifying its correctness. Prior work shows that review coverage and participation are associated with post-release quality [5], while perceived review quality depends on feedback thoroughness, reviewer familiarity, change quality, and actionable feedback [8, 13]. Reviewers rely on information about rationale, dependencies, expected behavior, surrounding context, and tests to evaluate changes [11, 39]. Missing or complex information can lead to confusion, delays, and increased review effort [36]. Reviewers therefore make integration decisions with incomplete information and must determine whether the available evidence is sufficient to approve a change.

Beyond technical evidence, reviewers also consider authorship, project history, and social context. Studies of pull-based development show that PR acceptance is associated with change size, test results, discussion activity, project membership, and contribution history [25, 26, 40]. Review latency is influenced by author-reviewer relationships, prior interactions, social connections, and workload [27, 41, 42]. Expertise and ownership provide additional context for review decisions. Developers associate review quality with reviewer expertise and code familiarity [8], reviewers develop subsystem ownership through repeated review activity [28], and participation is shaped by prior review and authoring experience [43]. Collectively, these studies show that conventional PR review draws on technical evidence, author experience, code ownership, and interaction history to inform integration decisions.

## 2.2 From AI Assistance to Agentic PRs

AI-assisted programming complicates this cue system because authorship is no longer binary. A PR may be human-written, AI-assisted, prompt-generated, manually revised after generation, or produced through an agentic workflow [18–20, 22]. Studies of Copilot and related tools show that developers use generation models to accelerate implementation and explore alternatives, but still need to understand, validate, and adapt generated code before relying on it [14, 18–20]. Recent work further shows that AI-code provenance affects validation and repair behavior [44], traces of AI tool use appear in commits and PRs [45], and PR descriptions can reduce reviewer uncertainty and improve outcomes [46].

We distinguish *AI assistance*, where AI supports a human-controlled process, from *AI authorship*, where an AI system is perceived to have produced a substantial part of the submitted change. An *HPR* is one where a human remains the primary decision-maker even if AI tools are used. An *AIPR* is produced substantially by an AI-based system, including coding agents [15–17] operating within repository workflows. This distinction matters because reviewers rarely observe exact line-level provenance [25, 45]. Instead, they infer authorship from PR descriptions, tests, commits, contributor behavior, and evidence of manual adaptation [11, 26, 44, 46]. In AIPRs, responsibility may be distributed across an AI system, human prompter, submitter, and future maintainers [19, 20], so conventional cues such as author track record, subsystem familiarity, and responsiveness may no longer carry the same meaning. Our study therefore focuses on *perceived authorship*: the reviewer's interpretation of whether a PR is primarily human- or AI-authored at review time, without assuming HPRs are free of AI assistance or AIPRs are free of human involvement [19, 22].

## 2.3 Human Review of Agent-authored PRs

Recent studies have examined how developers review contributions produced by AI agents. Repository studies report that many agent-generated PRs receive no human review and that human involvement often focuses on guiding or supervising agents rather than independently inspecting submitted code [47]. Analyses of review comments on agent-authored PRs identify concerns related to documentation, refactoring, coding style, testing, and security [48]. Human reviewers also engage in more review rounds for AI-generated code than for human-written code [49], while the descriptions agents provide for their changes are associated with review engagement and merge outcomes [50]. Practitioners have raised concerns about the growing imbalance between the effort required to generate contributions and the effort needed to review them [51, 52]. Repository evidence further associates AI-generated contributions with increased review pressure and defensive practices in open-source projects [53]. In response, projects have introduced policies that generally permit AI assistance while requiring disclosure, human involvement, and accountability [54–56]. Policy adoption has also been associated with greater AI-use disclosure and more extensive review interactions [57]. Although these studies provide evidence about review activities, outcomes, and project policies, they offer limited insight into how individual reviewers assess AI-authored changes. In particular, differences in initial trust, evidence requirements, and verification strategies between HPRs and AIPRs remain insufficiently understood. This study examines these differences from the reviewer's perspective.

### 2.4 Trust Calibration and Human Oversight

AI-generated code raises review-relevant concerns about correctness, robustness, performance, and security. Prior work shows that generated code may be sensitive to semantically equivalent prompts, contain insecure patterns, or fail when tasks require broader project contexts [30–32, 58]. Coding agents intensify this concern by moving AI support into repository workflows, where agent-authored PRs vary in task type, CI outcomes, review dynamics, and integration decisions [22–24, 38, 50, 59]. This work largely explains how developers use AI or how AIPRs appear in traces, but not how reviewers decide whether such PRs are trustworthy enough to approve. This gap connects AIPR review to trust calibration and appropriate reliance. Automation can be used, misused, disused, or abused [33], and reliance is appropriate only when trust is calibrated to capability, task context, and available evidence [34]. Trust is further shaped by risk, workload, prior experience, and interface cues [35]. Human-AI interaction research argues that systems should expose uncertainty and limitations so users can accept, question, or override AI output [60, 61], and that surfacing uncertainty can improve human judgment under automation [62]. In AIPR review, over-reliance may lead reviewers to approve plausible code without adequate scrutiny, while under-reliance may lead them to reject useful contributions despite sufficient evidence [33, 34]. Appropriate reliance therefore requires calibrating trust to code quality, change risk, evidence credibility, and accountability [34, 35]. Recent work also emphasizes the need to evaluate agentic development based on its ability to support trustworthy code integration, beyond code generation capabilities [63]. Practitioners expect human oversight to remain central for understanding, accountability, and organizational risk, and reviews of agent-authored PRs surface concerns about correctness, documentation, refactoring, testing, and security [48, 49, 64]. We use trust calibration to guide our comparison of initial trust, trust cues, reliance on evidence, and responses to uncertainty. We do not test a formal theoretical model or consider individual survey items validated measures of trust.

### 2.5 Trust and Provenance in the Evaluation of AI-Generated Code

Studies of developers' trust in AI code generation provide one foundation for understanding how generated contributions are evaluated. Wang et al. [65] interviewed 17 developers and used a subsequent design probe to identify difficulties with establishing appropriate expectations, configuring AI tools, and validating their suggestions. In a survey of 410 developers, Liang et al. [66] found that unmet functional and non-functional requirements and difficulty steering AI tools were important obstacles to adopting generated suggestions. Perry et al. [67] experimentally examined security-related programming tasks and found that participants with AI assistance produced less secure code while expressing greater confidence in its security. Khati et al. [68] synthesized research on trust in LLM-based software engineering and highlighted the need to distinguish subjective trust from system trustworthiness and to move beyond accuracy alone. These studies clarify why generated code may require validation, but primarily address developers using AI while producing code rather than reviewers judging a submitted contribution.

Related research further investigates how knowledge of code authorship influences reviewers' assessments. Huang et al. [69] controlled the apparent author of code-review tasks (man, woman, or machine) in a study combining behavioral measures with eye tracking and neuroimaging. Machine-labeled PRs were accepted less frequently, although the reported association was small and concerned an automated repair tool rather than a contemporary LLM agent. In a controlled study involving 28 participants, Tang et al. [44] found that disclosure of LLM provenance changed how developers inspected and repaired generated code, increasing some forms of validation effort and performance. More recently, Singh et al. [70] experimentally varied human-versus-AI provenance labels and code complexity for 385 developers deciding whether to accept an inline code revision. They found no statistically supported effect of the provenance label, whereas greater complexity increased acceptance of incorrect revisions. Thus, the evidence that provenance labels by themselves affect review outcomes is mixed and depends on the task and outcome being measured. These experiments manipulate source information under controlled conditions. Our study instead asks practitioners to compare their reported decision criteria across the broader HPR and AIPR contexts, including trust cues, sufficient evidence, uncertainty, and verification priorities.

Accountability and review workflow design offer an another perspective. Alami et al. [71] found that traditional peer review can transform individual responsibility into collective accountability, while introducing an LLM on the reviewer side challenges reciprocal accountability. Gullstrand Heander et al. [72] identified trust calibration as a central concern in reviewing LLM-generated multi-file changes and developed a layered, risk-aware review workflow through participatory design and evaluation. Repository evidence from Zhong et al. [49] further documents differences in human review interaction around generated code. Together, these studies suggest that authorship, evidence, and human responsibility cannot be reduced to a provenance label. For AIPRs, accountability may rest with the human prompter, submitter, or maintainer. The challenge for reviewers is determining whether that responsibility is clearly established and supported by credible validation evidence. Table 1 contrasts the designs and outcomes of these closely related studies with the decision criteria investigated in our study.

**Table 1** Comparison of studies examining trust, validation, and review of AI-generated code.

| Study | Design and sample | Authorship comparison | Outcome examined | Distinction from our study |
|---|---|---|---|---|
| Huang et al. [69] | Eye-tracking and neuroimaging study with 37 students | Assigned human and machine authorship labels, including male and female human authors | PR acceptance and reviewer attention | Examines responses to assigned labels on PRs from a pre-LLM automated repair tool |
| Tang et al. [44] | Laboratory study with 28 students | LLM provenance disclosed or undisclosed | Code validation and repair behavior | Examines provenance awareness while developers repair code rather than PR approval criteria |
| Singh et al. [70] | Between-subjects experiment with 385 developers | Human or AI authorship labels across complexity levels | Acceptance of correct and incorrect proposed revisions | Examines acceptance decisions for individual code revisions under assigned authorship labels |
| Perry et al. [67] | Security-focused programming experiment with 47 participants | AI-assistant access versus no access | Code security and participant confidence | Examines code creation and perceived security rather than PR review |
| Wang et al. [65] | 17 interviews followed by a design probe | No direct HPR/AIPR comparison | Trust in AI coding tools | Focuses on developers using AI coding tools rather than reviewing AI-authored contributions |
| Alami et al. [71] | 16 interviews and simulated review focus groups | Peer-led versus LLM-assisted reviewing | Perceived accountability and reciprocity | Examines AI assistance in reviewing rather than the assessment of AI-authored PRs |
| Gullstrand Heander et al. [72] | Participatory design with 17 practitioners and a follow-up survey of 43 | LLM-generated multi-file changes | Review requirements and design evaluation | Investigates review support for LLM-generated changes rather than comparing HPR and AIPR judgments |
| Zhong et al. [49] | Repository analysis of 278,790 review conversations | Observed interactions involving human- and AI-authored code | Review rounds and feedback characteristics | Examines observable review interactions without directly eliciting reviewers |
| **This study** | **Questionnaire survey with 239 practitioners** | **Within-respondent comparison of HPRs and AIPRs using matched items** | **Initial trust, trust cues, evidence requirements, uncertainty handling, and verification priorities** | **Examines how the same practitioners report their trust, evidence requirements, and verification priorities across HPR and AIPR contexts** |

## 2.6 Positioning of This Study

Prior work has examined conventional PR review, developers' use of AI coding tools, and the review of AIPRs in repositories. However, limited evidence exists on how reviewers' approval judgments differ between human- and AI-authored PRs. Controlled experiments examine the effects of authorship labels on specific decisions, while interview and design studies investigate trust and accountability without directly comparing HPRs and AIPRs (Section 2.5). We address this gap through a paired questionnaire survey that compares the same respondents' reported trust cues, evidence requirements, uncertainty handling, and verification priorities across HPR and AIPR contexts. Our study examines how reviewers assess whether the available evidence provides sufficient confidence to approve and merge AI-generated code.

# 3 Methodology

## 3.1 Research Questions

We use an online questionnaire survey with a within-respondent design, in which each respondent answered matched items about HPRs and AIPRs, to compare how reviewers evaluate the two across review decisions. Our study is guided by three Research Questions (RQs):

- **$RQ_1$: How does perceived authorship shape reviewers' initial trust and trust cue reliance in PR review?** This RQ focuses on whether reviewers begin with different trust assumptions for HPRs and AIPRs, and whether they rely on different cues when forming that trust. We compare author-centered cues for HPRs, AI-centered cues for AIPRs, and shared review cues such as change size, tests, and PR description clarity.
- **$RQ_2$: How does perceived authorship change what counts as sufficient evidence and how reviewers handle residual uncertainty before approval?** Here, we investigate whether AI authorship changes the evidence required before approval and how reviewers respond when parts of a change remain uncertain. We compare approval criteria, tolerance for unresolved uncertainty, and reported evidence thresholds across HPRs and AIPRs.
- **$RQ_3$: How does perceived authorship change reviewers' verification strategies and allocation of review effort?** This RQ examines whether AI authorship shifts review work toward defensive validation and accountability, including verification of non-trivial logic, review-time allocation, stop-review conditions, and up-front artifacts for reducing uncertainty.

## 3.2 Survey Design and Data Analysis Plan

To address the RQs, we conducted a cross-sectional mixed-method questionnaire survey. The survey design, recruitment, and analysis followed established guidance for empirical survey research in software engineering [73, 74] and the ACM SIGSOFT Empirical Standards [75]. A survey suits our RQs because a reviewer's initial trust, the cues behind it, the evidence the reviewer considers sufficient, and the action taken when uncertainty remains are rarely recorded in repository traces, which capture outcomes and comments rather than the reasoning behind them [76]. Each respondent answered matched questions about HPRs and AIPRs, allowing us to compare their reported review practices across both contexts within the same person. These comparisons reflect differences in reported judgments rather than causal effects of authorship on actual review behavior (Section 6.2). The questionnaire combined quantitative items for characterizing and comparing reviewer judgments with open-ended questions for examining the reasoning underlying those judgments.

### 3.2.1 Target Population and Sampling Objective

Our target population comprises software practitioners with code-review experience in development settings where AI-authored or agent-generated pull requests are relevant. Because no comprehensive sampling frame of such practitioners exists, we used complementary recruitment sources to reach reviewers with different levels of professional experience, project context, and exposure to AIPRs. In particular, recruitment included developers identified from repositories with known agent-authored PR activity, developers from mature and active open-source projects, and practitioners reached through professional networks.

Our sampling objective was to obtain a heterogeneous practitioner sample capable of providing experience-based evidence about AIPR review rather than to reproduce the demographic distribution of the broader software development population. Accordingly, the study examines patterns in reviewers' reported trust, evidence requirements, uncertainty handling, and verification practices. Because HPR and AIPR items were paired, we additionally examine within-respondent differences while holding constant the respondent's broader review experience and disposition. We characterize the achieved sample and its coverage boundaries explicitly and restrict our conclusions to the contexts supported by the recruited participants [75, 77].

### 3.2.2 Instrument Construction and Validation

We developed the questionnaire iteratively following guidance for survey research in software engineering [73, 74, 78, 79]. Candidate response options for closed-ended items were drawn from factors repeatedly identified in prior work and adapted to the AIPR setting. Trust cues drew on evidence that contributor track record, experience, project familiarity, change size, tests, and description quality shape PR evaluation [8, 25, 26, 28]. For AIPRs, we included additional cues, such as the human prompter's expertise and the AI tool's reputation, informed by research on developers' interactions with AI-generated code [19, 20]. Sufficient-evidence criteria and verification strategies drew on work on rationale, tests, and comprehension in review [10, 11, 36, 39], and up-front artifacts drew on work on PR descriptions and provenance [44, 46].

We then assessed the questionnaire through two stages of pre-deployment validation. First, two industry practitioners with 8+ and 17+ years of professional experience and two software-engineering researchers with 11+ and 6+ years of MCR experience independently reviewed the questionnaire for relevance, clarity, terminology, and coverage of the intended review scenarios. The University of Saskatchewan Behavioural Research Ethics Board also reviewed it for clarity and for leading, confusing, or double-barreled items. We revised the wording, response alternatives, and structure iteratively based on their feedback. Second, we conducted a pilot study with 11 software practitioners with code-review experience. The pilot was used to assess whether the questions and response alternatives were understandable in realistic review contexts, whether important response possibilities were missing, and whether the questionnaire flow and length were appropriate. Feedback from the pilot was used to refine ambiguous wording, revise or clarify response alternatives, and finalize the questionnaire before deployment. Responses from the pilot study were not included in the reported dataset.

The resulting questionnaire is a literature-informed instrument developed and refined through iterative validation, without relying on a previously established psychometric scale. Its items are single indicators of concrete judgments and actions rather than multi-item scales for latent attitudes [80], so we do not report scale reliability. Instead, we examine whether our interpretations are supported consistently across item formats and open-ended responses (Section 6.1) [81, 82]. Three design choices addressed known response biases. First, closed-ended items used item-specific response options, such as an ordered trust scale from *very low* to *very high* or a list of concrete review actions, rather than agree-disagree statements, which are more susceptible to acquiescence [83, 84]. Second, every paired item allowed respondents to report no difference, either by giving the same answer in both conditions or by selecting explicit options such as *no conscious calibration*, *about the same* evidence requirement, *usually continue reviewing*, or *none of these matter much*, so a stated contrast required an active choice. Third, every selection item included an *Other* or *It varies* option with free text, and a dedicated open-ended question asked respondents to explain their reasoning. The predefined response alternatives capture plausible reviewer considerations identified through prior research and pre-deployment validation, but they are not intended as an exhaustive taxonomy. Accordingly, we interpret option-level results as respondents' endorsement of the alternatives presented in the questionnaire, while the open-ended responses provide a complementary source of evidence for considerations outside the predefined response space. The paired blocks asked respondents to compare pull requests authored by human developers with those generated primarily by AI systems, consistent with the definitions in Section 2.2. The complete questionnaire is provided in the replication package.

#### 3.2.3 Questionnaire Structure and Measures

The final questionnaire combined closed-ended, multiple-selection, ordinal-rating, paired-comparison, and open-ended questions. We used these formats for complementary purposes: structured items support systematic comparison across respondents and conditions, whereas open-ended questions allow respondents to explain their reasoning and introduce considerations not represented by the predefined alternatives. Table 2 summarizes the questionnaire blocks analyzed in this paper, their response formats, and their relationship to the research questions. The complete questionnaire, including the original wording, response alternatives, and item order, is provided in the supplementary material.

The paired blocks follow the course of a review decision. The trust blocks concern the assumptions and cues reviewers bring to a PR before detailed inspection. The verification and time-allocation blocks concern how reviewers establish confidence during review, whereas the sufficient-evidence and unresolved-uncertainty blocks concern the point of approval. The stop-review and up-front artifact blocks examine when reviewers pause and what information they want before investing further effort. The comparative block then asks respondents directly whether their evidence requirement for AIPRs differs from that for HPRs, which escaped defect feels more like reviewer failure, and what makes deciding whether an AIPR is safe to approve more difficult. Open-text options throughout the questionnaire and this open-ended question provide complementary qualitative evidence about the reasoning behind respondents' choices. Most response options were identical across the two versions. Options that refer to the person responsible for the change were worded for each condition, and we refer to them as parallel options: the author's familiarity with the subsystem versus the human prompter's familiarity with the subsystem as a trust cue, the author's explanation versus the human prompter's explanation or PR description as an approval criterion, requesting clarification from the author versus from the human prompter, and coordinating with owners versus with the human prompter. The trust shortcut was phrased as trusting the author's reasoning for HPRs and as trusting the output if tests pass for AIPRs. Identical and parallel options are compared with paired tests, and options that apply to only one condition, such as the author's track record or the AI tool's reputation, are reported descriptively (Section 3.2.6).

#### 3.2.4 Recruitment and Sampling Frame

We used a purposive, multi-frame recruitment strategy to reach practitioners with relevant experience in code review. First, we targeted reviewers from the AIDev dataset [23], which contains repositories with PRs created by coding agents including Claude, Codex, Devin, and Copilot. We retained projects with at least 500 stars and identified developers who had posted at least one review comment, yielding 516 publicly available email addresses. Second, we broadened the repository-based sampling frame beyond projects with known AIPR activity. We selected 133 active Apache Software Foundation projects and 49 additional mature GitHub projects using project activity, popularity, contributor count, commit history, and presence of active PRs as selection criteria. The selected projects covered several widely used programming languages, including Java, Python, C, C++, and JavaScript, and we verified that projects did not overlap across the repository sets. This broader repository-based process identified more than 4,000 eligible developers, of whom 3,357 had publicly available email addresses. Detailed project-selection criteria are reported in the supplementary material.

**Table 2** Structure of the questionnaire used in this study.

| Survey block | Question type | Purpose | RQ |
|---|---|---|---|
| Participant background | Closed-ended and demographic | Characterize country, software-development experience, code-review experience, and self-reported exposure to AI-authored or agent-generated PRs. | Context |
| Initial trust | Paired ordinal ratings | Compare, within the same respondent, initial trust in the correctness of code submitted in HPRs and AIPRs. | $RQ_1$ |
| Factors influencing trust | Paired multiple-selection | Identify the author-centered, AI-centered, and shared cues that influence respondents' initial trust in HPRs and AIPRs. | $RQ_1$ |
| Verification of non-trivial logic | Paired multiple-selection | Compare the strategies respondents use to verify the correctness of non-trivial logic in HPRs and AIPRs. | $RQ_3$ |
| Sufficient evidence for approval | Paired single-choice | Compare what most often makes respondents decide that they have verified a change enough to approve. | $RQ_2$ |
| Unresolved uncertainty | Paired single-choice | Compare respondents' most typical action when part of a change remains uncertain and cannot be verified within time. | $RQ_2$ |
| Evidence requirement comparison | Ordinal, categorical, and open-ended | Elicit respondents' evidence requirement for AIPRs relative to HPRs, which escaped defect feels more like reviewer failure, and what makes AIPR approval more difficult. | $RQ_2$ |
| Review time allocation | Paired rankings | Compare how respondents rank eight review activities by time spent when reviewing merge-candidate HPRs and AIPRs. | $RQ_3$ |
| Stop-review conditions | Paired top-two selection | Compare the situations that most often make respondents stop reviewing and require more information before proceeding. | $RQ_3$ |
| Up-front artifacts | Paired top-two selection | Compare the up-front artifacts respondents consider most useful for efficient review of HPRs and AIPRs. | $RQ_3$ |

Finally, we disseminated the survey through LinkedIn to broaden participation beyond developers identified through public repository histories. Combining repository-derived recruitment with professional-network dissemination was intended to reduce dependence on a single sampling frame and to capture practitioners working across different development contexts. The survey was administered through SurveyMonkey[1].

### 3.2.5 Response Screening and Final Sample

The survey received 261 response records during the two-month data-collection period. Before analysis, we screened the records for consent and questionnaire completion. We excluded one response for which consent was not provided and 21 substantially incomplete responses, resulting in a final dataset of 239 participants. We classified a response as substantially incomplete when more than 50% of the questions were unanswered. Optional and open-ended questions were analyzed using the number of valid responses available for each item, and we report item-specific denominators when they differ from the full sample.

[1] https://www.surveymonkey.com/

Since recruitment combined direct email invitations with open dissemination through LinkedIn, the total number of individuals who encountered the survey is not observable. We therefore report the observed response disposition instead of a single invitation-level response rate.

### 3.2.6 Quantitative Analysis

We matched the quantitative analysis to the measurement scale and dependency structure of each survey item. Descriptive statistics are the primary evidence for single-condition questions: for categorical and multiple-selection items, we report valid counts and percentages, and for ordinal items we report response distributions and summary statistics appropriate to ordered data.

For questions that directly compared HPR and AIPR conditions within the same respondent, we treated observations as paired. Initial trust ratings and the ranks of review activities were analyzed using Wilcoxon signed-rank tests, with matched-pairs rank-biserial correlation reported as an effect-size measure [85, 86]. We additionally report the proportion of respondents whose AIPR response was lower than, equal to, or higher than the corresponding HPR response to make the direction of within-participant change transparent. For paired single-choice questions with more than two categories, namely sufficient-evidence criteria and actions under unresolved uncertainty, we used Stuart–Maxwell tests of marginal homogeneity [87, 88], followed by exact McNemar tests for individual categories [89]. To summarize whether uncertainty handling became stricter, we ordered the actions from approving and documenting the uncertainty to blocking approval and confirmed that the result holds under alternative orderings. For multiple-selection and top-two questions, each response alternative was represented as a binary indicator. Although respondents were instructed to select their top two options, the survey interface did not enforce this limit. We therefore analyzed these items as multiple-selection responses rather than strict top-two rankings. Alternatives shared by both conditions were compared using exact McNemar tests [89, 90], and alternatives specific to one condition were reported descriptively. We treat these rates as endorsement of the response alternatives presented in the questionnaire rather than as an exhaustive ranking of all possible reviewer considerations. Similarly, because each respondent ranked all eight review activities, we interpret ranks as relative time priorities. For the rankings within each condition, we used the Friedman test [91] with Kendall's $W$ as an effect-size measure [92] to assess agreement on priorities, and for the comparative evidence-requirement item, we used a one-sample Wilcoxon signed-rank test against the neutral midpoint.

Exploratory analyses examined whether selected patterns varied with respondents' AIPR exposure, software-development experience, or code-review experience. Depending on the outcome type, we used Spearman's rank correlation [93], Kruskal–Wallis tests [94], or logistic regression [95], with Benjamini–Hochberg correction where many cue-level models were fitted. These analyses are treated as contextual rather than confirmatory. In addition, several analyses involved multiple related comparisons addressing the same construct: shared trust cues, verification strategies, sufficient-evidence criteria, uncertainty actions, review activities, stop-review conditions, and up-front artifacts. Testing each comparison independently at $\alpha = .05$ would increase the probability of at least one false-positive result within such a family. We therefore controlled the family-wise error rate within each set of related comparisons using the Holm procedure [96]. Holm's sequential procedure provides strong control of the family-wise error rate while being less conservative than the standard Bonferroni correction. All tests were two-sided with $\alpha = .05$. We interpret statistical significance together with effect sizes and response distributions rather than relying on $p$-values alone.

### 3.2.7 Qualitative Analysis

We analyzed the open-ended responses using codebook-based thematic coding [97]. We use this term to distinguish our procedure from reflexive thematic analysis, because our analysis employed an explicit codebook, independent coders, and inter-rater agreement.

Two researchers with 10+ and 8+ years of software-development experience first independently examined a random sample of 100 responses. They identified recurring concepts, compared the independently generated categories, and iteratively consolidated them into a shared codebook using a card-sorting procedure [98]. Categories were defined with descriptions and decision rules, and responses could receive multiple codes when they expressed more than one concept. After stabilizing the codebook, the two coders independently applied it to the remaining responses. For open-ended items with fewer than 20 valid responses, both coders examined all available responses. We measured inter-rater agreement on independently coded samples using Cohen's $\kappa$ [99]. Disagreements were resolved through discussion, and cases that remained unresolved were adjudicated by a third evaluator with more than 20 years of software-engineering experience.

The final codes were subsequently grouped into higher-level themes where conceptually related categories described the same broader review phenomenon. Because multi-label coding was permitted, theme percentages are not mutually exclusive. We use the qualitative responses to explain, contextualize, and occasionally qualify the quantitative patterns rather than to estimate population prevalence.

# 4 Evaluation and Results

## 4.1 Overview of Participant Demographics

Over two months, the survey received 261 responses. After excluding 22 responses, one without consent and 21 substantially incomplete, we analyzed 239 valid responses. Participants came from 31 countries, with the largest groups from the United States (19.7%, 47), Canada (15.5%, 37), Bangladesh (10.9%, 26), Germany (10.9%, 26), and India (7.1%, 17). The top five countries accounted for 64.0% of responses, while the top 10 accounted for 83.7%, indicating geographic breadth with some concentration in major respondent countries.

Participants varied in professional experience. Most had substantial software development experience, with 58.2% (139) reporting more than seven years and 30.1% (72) at least 16 years. Code review experience was also broad, with 48.5% (116) reporting more than seven years and 32.6% (78) reporting 0–3 years. The sample therefore includes many experienced developers and reviewers while retaining early-career perspectives. Participants also reported meaningful exposure to AIPRs: 78.2% (187) had reviewed at least 10 AIPRs, 62.8% (150) more than 20, and 49.4% (118) more than 40. Overall, the sample is grounded in practical AIPR review experience while still including respondents with more limited exposure. Figure 1 shows the full respondent distributions.

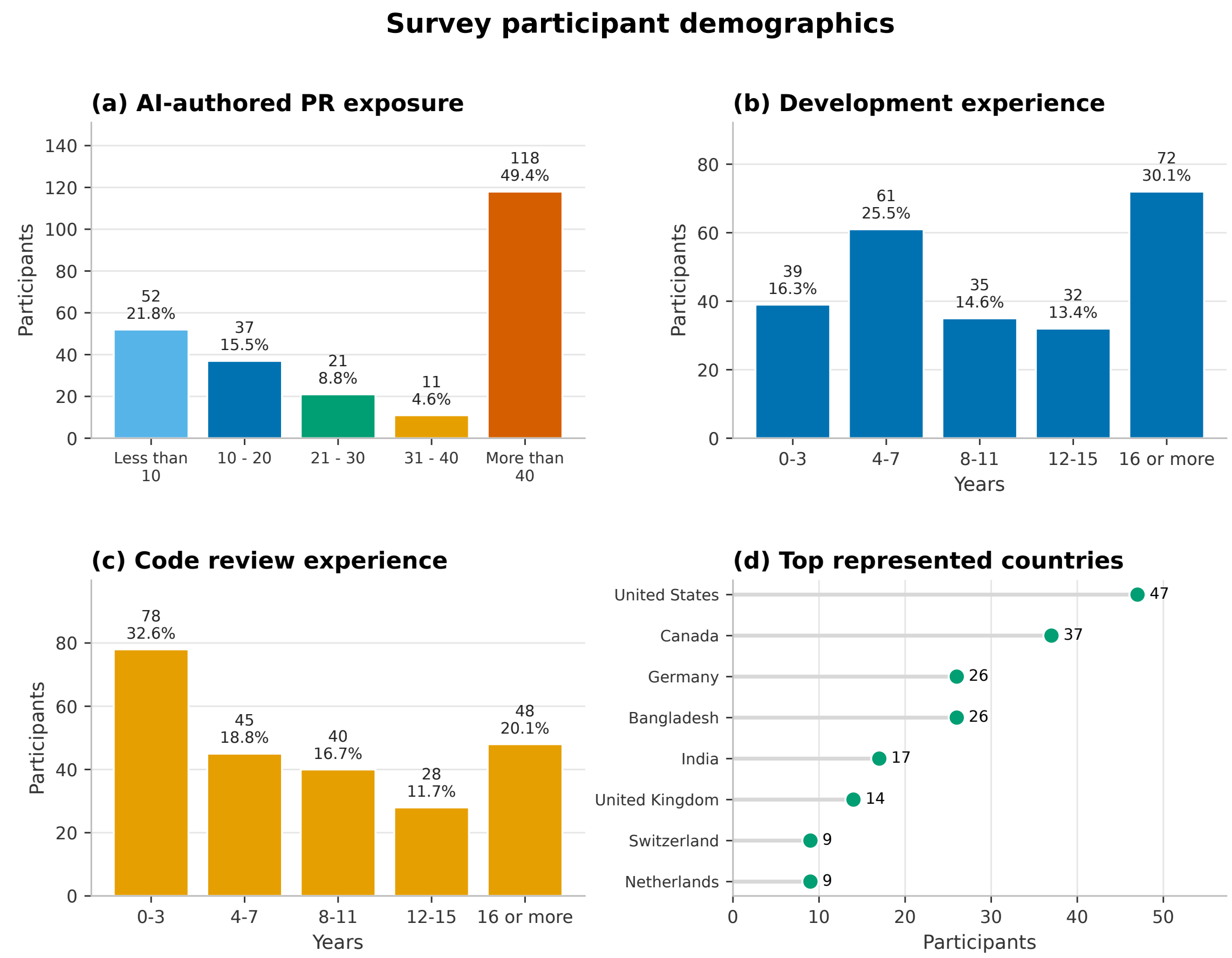

**Fig. 1** Participant demographics. (a) shows participants' self-reported exposure to reviewing AI-authored or agent-generated PRs; (b) and (c) summarize software development and code review experience, respectively; (d) shows the eight most represented countries in the analyzed sample. Percentages are computed over the 239 valid responses.

## 4.2 $RQ_1$: Authorship and Trust Calibration

$RQ_1$ examines whether perceived authorship changes reviewers' initial trust and the cues they use to calibrate that trust. We first compare HPR and AIPR trust ratings, then analyze the selected trust cues.

### 4.2.1 Initial trust

Reviewers reported substantially lower initial trust in AIPRs than in HPRs. For HPRs, 52.7% selected *high* or *very high* initial trust and only 6.7% selected *low* or *very low*. For AIPRs, these shares shifted to 15.5% and 40.2%, respectively. At the paired level, 64.9% rated AIPRs lower than HPRs, 24.3% reported the same trust level, and 10.9% reported higher trust for AIPRs. A Wilcoxon signed-rank test confirmed significantly lower initial trust for AIPRs ($p < .001$), with a large matched-pairs rank-biserial effect ($r_{rb} = -0.77$, 95% CI $[-0.86, -0.68]$). Figure 2 shows the full paired trust distribution.

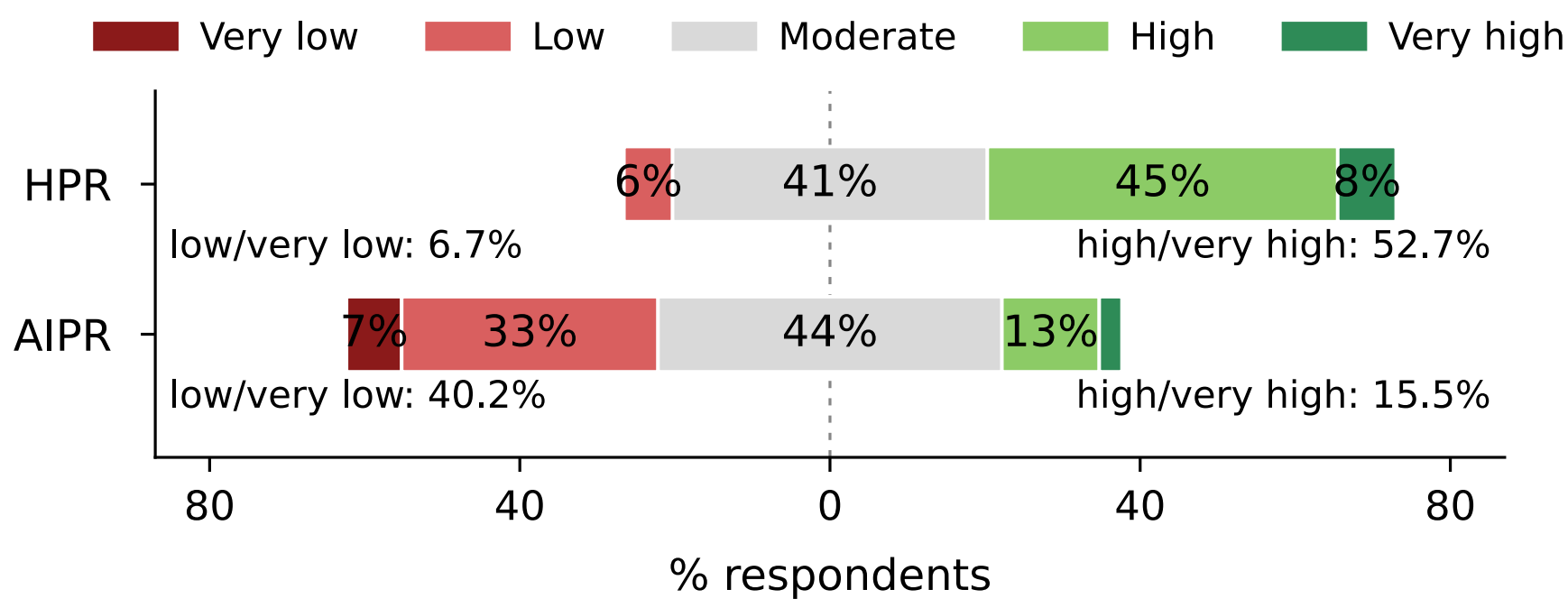

**Fig. 2** Initial trust calibration by perceived authorship.

We further examined whether the trust penalty was concentrated in particular respondent groups. We define the *trust penalty* as the human-authored trust score minus the AI-authored trust score, so positive values indicate lower initial trust for AIPRs. The median penalty was 1.0 in every background group, meaning that the typical respondent rated AIPRs one scale point lower than HPRs. AIPR review exposure showed only a weak negative association with the penalty ($\rho = -0.13$, $p = .05$), suggesting a slightly smaller trust gap among respondents with more AIPR exposure. However, this pattern was not significant across exposure groups in a Kruskal–Wallis comparison ($p = .32$) and was only marginal in logistic regression predicting whether a respondent rated AIPRs lower than HPRs (OR = 0.85, $p = .078$). Software development and code review experience were not significant predictors. Thus, lower initial trust in AIPRs appears broadly distributed rather than driven by less experienced reviewers. Detailed paired-shift distributions and the full HPR–AIPR transition matrix are reported in the supplementary material.

**Finding 1.1:** Respondents reported substantially lower initial trust in AIPRs than in HPRs. 64.9% rated AIPRs lower, with a median one-point drop and a large paired effect ($r_{rb} = -0.77$), although 35.2% reported the same or higher trust.

### 4.2.2 Trust cues

Having established lower initial trust in AIPRs, we examined which cues reviewers use to calibrate that trust. Respondents selected all cues influencing their initial trust for HPRs and AIPRs. We coded each cue as selected or not selected, reporting condition-specific cues descriptively and comparing shared cues with paired tests [89, 96].

Respondents described a change not only in the level of trust, but also in the basis on which trust is formed. For HPRs, respondents rely heavily on author-centered cues, especially the *author's prior track record* (77.9%) and *author's experience level* (59.6%). For AIPRs, the most selected condition-specific cue is *human prompter's expertise* (57.5%), while *AI tool reputation* is selected by 32.3%. This suggests that reviewers weigh accountable human oversight more than the reputation of the AI tool itself. Figure 3 shows the paired results for shared cues. The largest shift concerns *subsystem familiarity*, which decreases from 72.5% for HPRs to 44.1% for AIPRs (−28.4 pp, Holm-adjusted $p < .001$). At the paired level, this decrease is driven by respondents who select *subsystem familiarity* only for HPRs (82 versus 19 in the opposite direction), suggesting that subsystem knowledge carries less weight when the contributor is perceived as an AI system rather than a human with project experience. Because the HPR version of this option referred to the author's familiarity with the subsystem and the AIPR version to the human prompter's familiarity (Section 3.2.3), the decrease suggests that reviewers give less weight to a prompter's knowledge of the subsystem than to an author's. Part of this shift may therefore reflect the wording rather than authorship alone. In contrast, *change size* becomes more salient for AIPRs, increasing from 45.0% to 57.7% (+12.6 pp, Holm-adjusted $p = .005$). Other shared cues, including *change complexity*, *presence of tests*, *PR description clarity*, *release/bug-fix urgency*, and *no conscious calibration*, do not differ significantly after correction. Respondents also selected fewer cues for AIPRs than for HPRs (mean 3.76 versus 4.33, Wilcoxon $p < .001$, $r_{rb} = -0.47$), with 48.2% selecting fewer cues for AIPRs and 22.1% selecting more, which suggests that AIPR trust rests on a narrower set of signals.

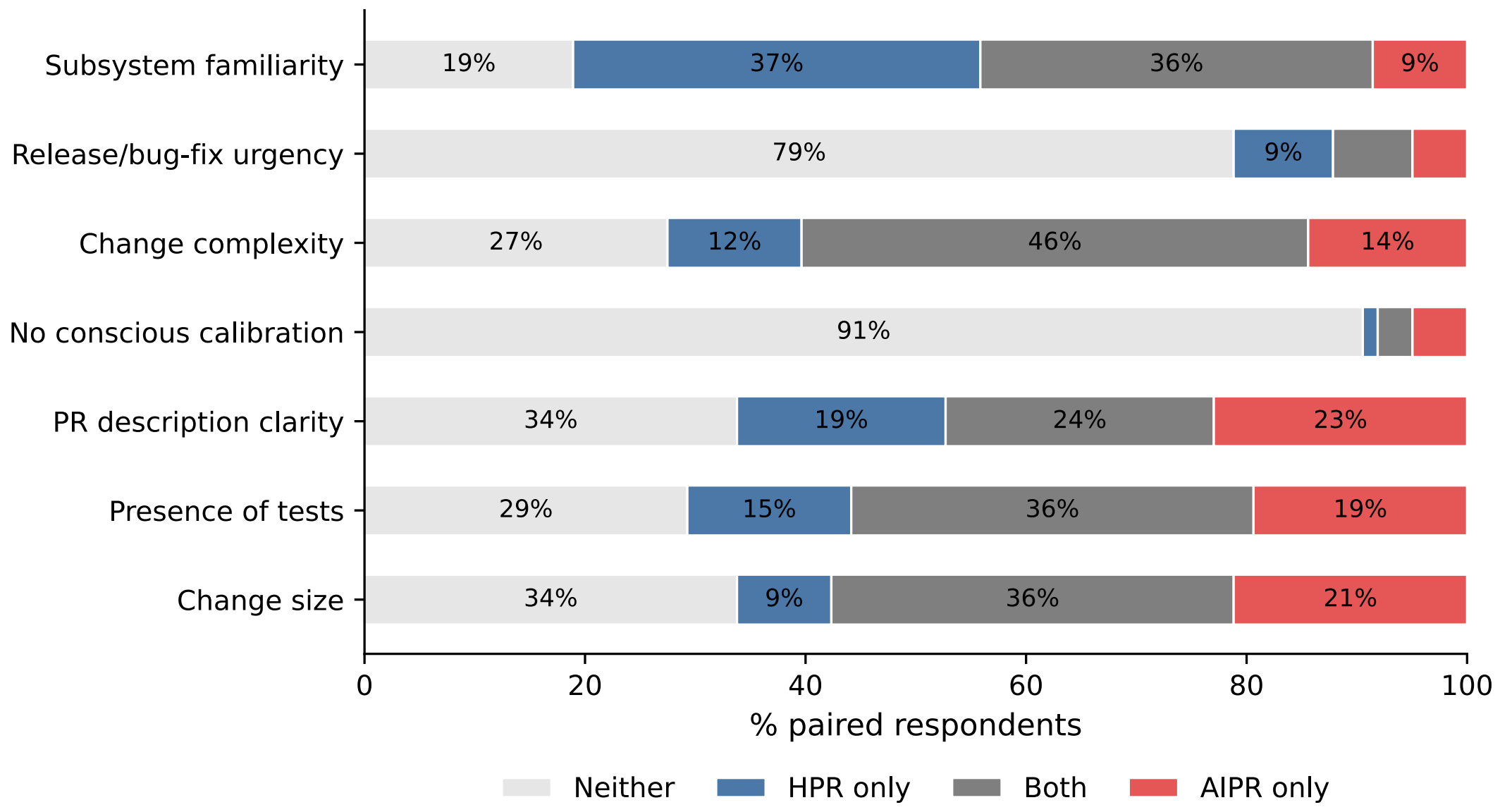

**Fig. 3** Selection of shared trust cues in HPRs and AIPRs.

The optional free-text responses were used as qualitative context rather than a standalone thematic result. Five HPR responses mainly elaborated conventional cues, such as visible code quality and fit with existing code flow, while the 17 AIPR responses more often named process and ownership cues, including codebase adaptation, human prompter understanding, and the extent of AI generation.

**Finding 1.2:** Respondents' HPR trust cues centered on author track record, experience, and subsystem familiarity, whereas their AIPR trust cues shifted toward human prompter expertise, bounded change size, and accountable oversight, and were fewer in number.

**Answer for RQ$_1$:** Respondents described trust calibration under perceived AI authorship as starting from lower confidence and drawing on a narrower set of signals treated as credible assurance. They no longer relied on conventional contributor-based cues in the same way. Instead, they looked for evidence that the AI-authored contribution is reviewable, supervised, and tied to accountable human judgment. Thus, AIPR trust is less about replacing the human author with the AI tool, and more about re-establishing accountability around a generated change.

## 4.3 RQ$_2$: Evidence and Uncertainty

This RQ examines whether perceived authorship changes reviewers' approval thresholds and tolerance for residual uncertainty. Building on the lower initial trust observed in RQ$_1$, we analyze what reviewers consider sufficient evidence, how they act when part of a change remains uncertain, and whether they report higher evidence requirements for AIPRs.

### 4.3.1 Sufficient evidence for approval

We first asked respondents what most often makes them decide that they have verified an HPR or AIPR "enough" to approve. Because these are paired nominal responses, we tested whether the overall distribution of sufficient-evidence criteria differed across authorship conditions using the Stuart–Maxwell test, followed by Holm-corrected McNemar tests for individual criteria.

Figure 4 shows that sufficient-evidence criteria differed significantly between HPRs and AIPRs ($\chi^2 = 48.71$, $df = 4$, $p < .001$). For HPRs, the most common approval criterion was tests covering the risky behavior (41.0%), followed by the reviewer being able to reason through all critical cases (28.5%). For AIPRs, this ordering reversed: reasoning through critical cases became dominant (55.2%), while reliance on tests dropped to 23.8%. Post-hoc paired tests confirm these shifts. Personal reasoning increased by 26.8 pp for AIPRs (28.5% to 55.2%, Holm-adjusted $p < .001$), whereas test coverage decreased by 17.2 pp (41.0% to 23.8%, Holm-adjusted $p < .001$). Explanation-based assurance, worded as the author's explanation for HPRs and as the human prompter's explanation or PR description for AIPRs, also decreased from 15.9% to 8.8%, but not significantly after correction ($p = .074$). Overall, 49.8% of respondents changed their sufficient-evidence criterion between HPRs and AIPRs, and 50.2% kept the same criterion.

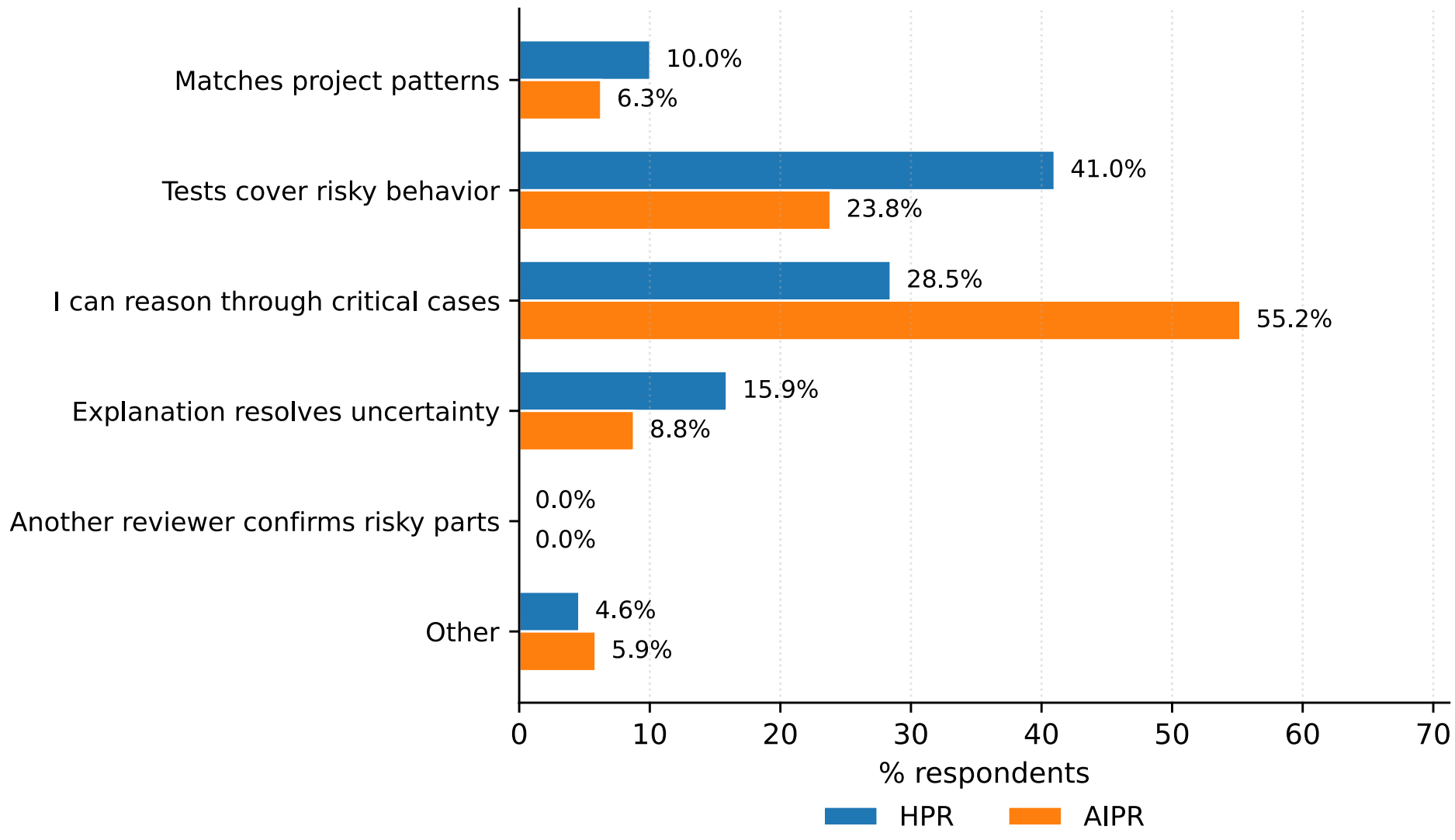

**Fig. 4** Sufficient-evidence criteria for approving HPRs and AIPRs.

The largest respondent-level shifts further clarify the pattern: 44 respondents (18.4%) moved from tests for HPRs to personal reasoning for AIPRs, and 24 (10.0%) moved from explanation-based assurance to personal reasoning. Movements in the opposite direction were rare: six respondents moved from personal reasoning to tests and three from personal reasoning to explanation. Thus, AIPR review does not simply increase demand for more tests. It shifts approval toward reviewer-owned reasoning about critical cases. Optional open-ended responses were sparse, with 11 HPR responses (4.6%) and 14 AIPR responses (5.9%), so we used them as qualitative context. HPR responses mainly combined conventional criteria such as low merge risk, tests, and author explanations. AIPR responses raised process concerns, including review bandwidth for agentic PRs, superficial AI-generated tests, scope creep, and explanations that increased rather than reduced uncertainty.

**Finding 2.1:** Under AI authorship, respondents' stated approval threshold moved from evidence presence to reviewer-owned assurance. AIPRs were less often judged sufficient based on tests and more often required reviewers to establish confidence through critical-case reasoning.

### 4.3.2 Residual uncertainty before approval

We next examined whether perceived authorship changes reviewers' tolerance for residual uncertainty. Respondents reported what they would typically do when one part of a change seemed probably not wrong but could not be verified within time, for both conditions.

Figure 5 shows a clear shift toward stronger review control under AI authorship. The overall paired distribution differed significantly between HPRs and AIPRs ($\chi^2 = 68.57$, $df = 4$, $p < .001$). For HPRs, the most common response was to request targeted evidence before approval (44.7%), followed by approving while documenting the uncertainty (19.8%) and blocking until resolution (19.4%). For AIPRs, blocking became dominant (43.9%), while approval with documented uncertainty dropped to 5.5% and targeted evidence requests to 28.7%. Post-hoc paired tests confirm this escalation. Compared with HPRs, AIPRs were less often handled by approving with documented uncertainty (−14.3 pp, Holm-adjusted $p < .001$) or requesting targeted evidence (−16.0 pp, $p < .001$), and more often blocked until resolution (+24.5 pp, $p < .001$). Asking another reviewer increased from 11.0% to 16.5%, but not significantly after correction ($p = .107$). Overall, 51.9% of respondents changed their uncertainty-handling action between HPRs and AIPRs. Among responses that could be ordered by review-control strictness, 41.4% became stricter for AIPRs, 11.7% became less strict, and 46.8% remained unchanged (sign test $p < .001$, Wilcoxon $p < .001$, $r_{rb} = 0.63$).

The open-ended *"It varies substantially"* responses provide qualitative context. This option was selected by 12 HPR and 13 AIPR respondents. HPR explanations mainly referred to risk-sensitive tolerance, including change criticality, product impact, rollback support, team discussion, or pairing with the developer. AIPR explanations included similar risk reasoning, but added AI-specific safeguards such as asking the human prompter to explain the uncertain part, splitting or redoing the PR, or rejecting AI-authored PRs. Thus, residual uncertainty in AIPRs appears less likely to be treated as a manageable review note and more likely to trigger blocking, rework, or explicit human accountability.

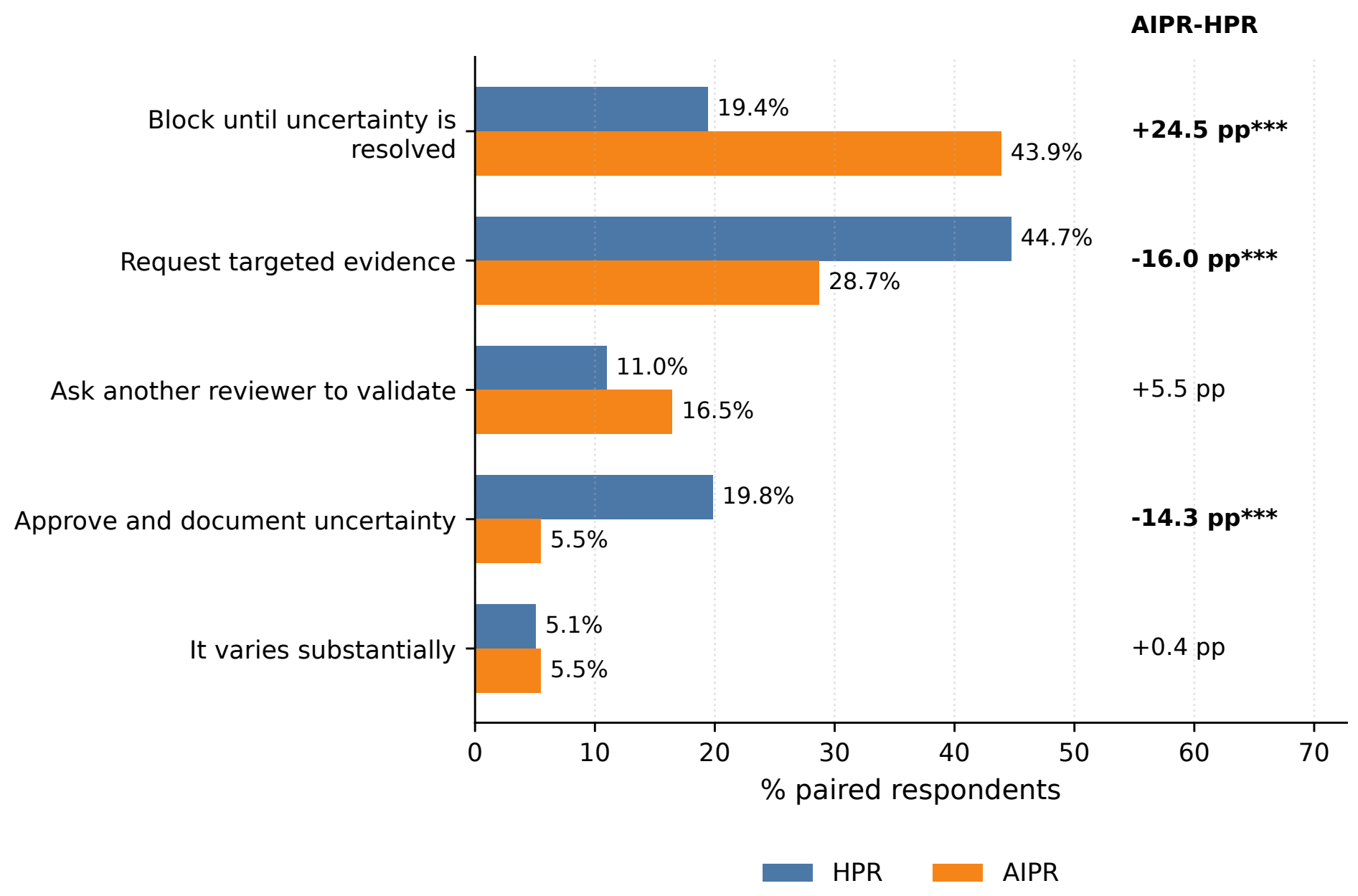

**Fig. 5** Reviewer actions at residual uncertainty. Asterisks mark differences that remain significant after Holm correction (exact McNemar tests): ** $p < .01$, *** $p < .001$.

**Finding 2.2:** Respondents reported less tolerance of residual uncertainty under AI authorship. Compared with HPRs, AIPRs were much less often approved with documented uncertainty and much more often blocked until the uncertainty was resolved.

### 4.3.3 Stated evidence requirements and rationales

The preceding analyses show that AIPRs change both approval criteria and uncertainty handling. We next tested whether respondents explicitly report a higher evidentiary threshold for AIPRs. This item offered a balanced five-point scale from *lower* to *higher* with *about the same* as midpoint. A majority reported a higher threshold: 61.1% selected *slightly higher* or *higher*, compared with 27.6% selecting *about the same* and 11.3% selecting *lower* or *slightly lower* requirements. The median response was *slightly higher*. A one-sample Wilcoxon signed-rank test against the neutral midpoint confirmed a significant shift toward higher evidence requirements ($W = 2269.0$, $p < .001$, $r_{rb} = 0.70$).

To explain this threshold, we thematically coded open-ended responses ($\kappa = 0.91$) to an open-ended question about what makes AIPR approval more difficult than HPR approval. Among 225 respondents, 158 (70.2%) described AIPR approval as harder. Table 3 summarizes the eight themes. Responses show that this threshold is not only general skepticism toward AI. Reviewers identify specific weaknesses in the evidentiary basis of AIPRs: generated code can appear plausible while hiding subtle errors, miss project-specific constraints, obscure intent and human understanding, and provide tests or explanations that themselves require verification.

**Finding 2.3:** Respondents reported a higher evidentiary burden for AIPRs and, among those who described AIPR approval as harder, attributed it to doubts about both the generated code and the credibility of its supporting evidence, including explanations, context fit, and accountable human understanding.

**Answer for RQ$_2$:** Respondents described a higher evidentiary bar under perceived AI authorship, grounded in a change in what they treat as credible assurance. For AIPRs, they looked for support that is project-grounded, independently verifiable, and tied to a responsible human. When that support was incomplete, they reported treating the gap as a reason to delay or block merging rather than as acceptable review risk.

**Table 3** Themes explaining why AIPR approval is perceived as more difficult than HPR approval. Percentages are over the 158 responses that described AIPR approval as harder and a response can carry several themes.

| Theme | Context | Example response |
|---|---|---|
| Plausible-but-wrong correctness (n = 46, 29.1%) | AI-authored code may look clean, idiomatic, or convincing while still hiding subtle logic errors, hallucinated behavior, missed edge cases, or incorrect assumptions. | *"AI-generated code often looks correct but in fact is not. It may also not be idiomatic, for the language or the project."* |
| Missing project context and domain constraints (n = 64, 40.5%) | Reviewers must verify whether the change fits broader codebase context, business logic, architecture, undocumented constraints, existing patterns, and downstream effects. | *"AI usually only sees a small portion of the system and doesn't understand the full application, previous technical decisions, or existing shared components."* |
| Unclear intent, rationale, and human understanding (n = 47, 29.7%) | Reviewers cannot rely on human intent in the same way; they may need to reconstruct why the change exists and whether the human submitter actually understands it. | *"With AI-generated code there's no author intent to appeal to, so if something looks off I have to figure it out myself."* |
| Weak or misleading evidence and tests (n = 25, 15.8%) | Tests, comments, PR descriptions, or explanations may be shallow, overconfident, excessive, or misaligned with the risky behavior that needs validation. | *"I don't trust LLM-generated evidence, only code. LLMs tend to be over optimistic."* |
| Large scope, volume, and review overload (n = 42, 26.6%) | AI-authored PRs may touch more files, add more code, or create more sprawling changes, increasing reviewer fatigue and making critical changes harder to isolate. | *"AIs can change large amounts of code making it easier to get overwhelmed by sheer volume."* |
| Unnecessary complexity, duplication, and maintainability risk (n = 43, 27.2%) | AI may over-engineer, duplicate existing functionality, introduce unfamiliar dependencies or patterns, or produce code that works locally but is harder to maintain. | *"AI tends to cause a lot of duplication so it is harder to review from within the small context of just files changed."* |
| Ownership, accountability, and communication gap (n = 47, 29.7%) | Reviewers may lack an accountable author who can explain, defend, maintain, or answer follow-up questions about the implementation. | *"Accountability is a major factor. With a human-authored pull request, I can usually discuss the implementation with the author."* |
| Security, safety, and compliance concerns (n = 15, 9.5%) | Some respondents emphasized vulnerabilities, sensitive data exposure, credentials, licensing/IP concerns, malware, or safety-critical risk. | *"Sometimes AI codes are highly suspicious than human authored code, such as exposing credentials, security vulnerabilities, or environment variables."* |

## 4.4 $RQ_3$: Verification and Effort

$RQ_3$ investigates whether perceived authorship changes verification strategies and review effort. Building on the trust and evidence shifts reported in $RQ_1$ and $RQ_2$, we organize the analysis around four aspects of review work.

### 4.4.1 Verification of non-trivial logic

We first examine verification of non-trivial logic, where reviewers must judge behavioral correctness rather than surface plausibility. Respondents selected all strategies they use for HPRs and AIPRs.

Figure 6 shows that AI authorship does not simply increase the number of verification strategies. The average number of active strategies is similar for HPRs and AIPRs (3.03 vs. 2.90, Wilcoxon $p = .155$, $r_{rb} = -0.14$). At the paired level, 21.8% of respondents select more active strategies for AIPRs, 48.7% select the same number, and 29.5% select fewer. Thus, the difference lies less in verification breadth than in strategy emphasis. At the strategy level, AIPR review shifts away from trust-based and explanation-based verification. The largest decrease concerns the trust shortcut: 52.6% select trusting the author's reasoning for HPRs, whereas only 15.0% select trusting AI output if tests pass for AIPRs (−37.6 pp, Holm-adjusted $p < .001$). Requesting clarification also decreases from 70.9% to 50.0% (−20.9 pp, $p < .001$), suggesting that clarification from a human prompter does not fully substitute for clarification from a human author. Mental simulation also decreases from 56.4% to 46.2% (−10.3 pp, $p < .001$). In contrast, concrete validation becomes more common for AIPRs. Running or inspecting tests increases from 52.1% to 65.0% (+12.8 pp, $p < .001$), and re-implementing or partially rewriting logic increases from 11.1% to 20.5% (+9.4 pp, $p < .001$). Cross-checking requirements and examining related files or history remain common in both conditions, but do not differ significantly after correction. In addition, 50.9% of paired respondents select the AIPR-specific option of applying stricter scrutiny than for human-authored code.

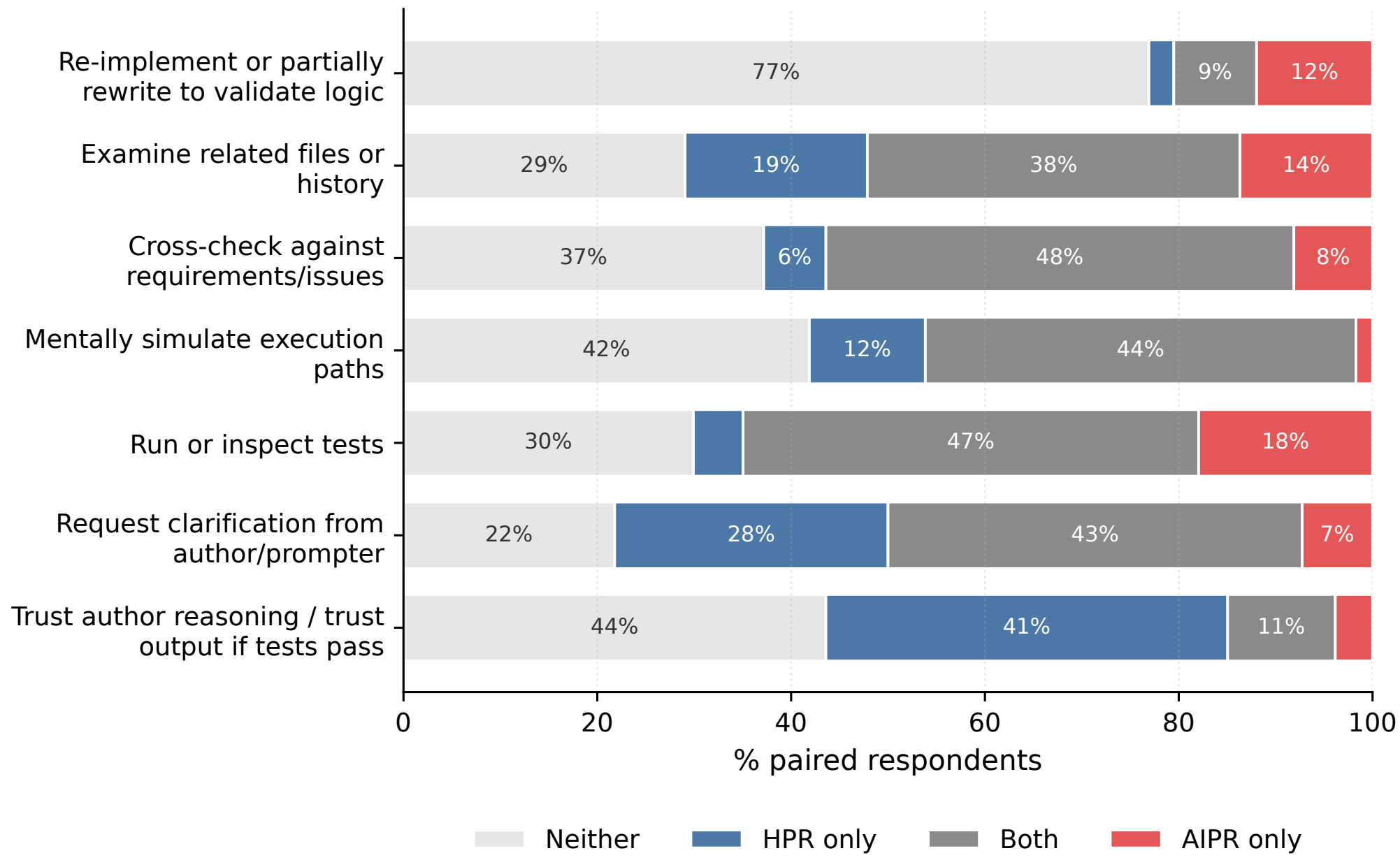

**Fig. 6** Paired selection states for verification strategies at non-trivial logic.

Optional open-ended responses were sparse, with 9 HPR responses (3.8%) and 15 AIPR responses (6.4%). HPR responses mainly describe conventional validation, such as local execution, CI inspection, additional tests, or discussing alternatives with the author. AIPR responses include similar checks but add AI-specific risk controls, such as asking agents to explain suspicious logic, using AI to analyze the implementation or design tests, splitting AI-generated PRs into smaller reviewable changes, strengthening project-specific lint rules, avoiding low-value review effort, or rejecting the PR until a human explains or implements the change. Thus, AIPR verification is not broader overall, but more oriented toward defensive validation, risk containment, and human accountability.

**Finding 3.1:** For non-trivial logic, respondents described a change in verification profile rather than verification breadth under AI authorship. They moved away from trust shortcuts and explanation-based checks toward concrete validation, defensive scrutiny, risk containment, and explicit human accountability.

### 4.4.2 Review time allocation

We next investigated whether the changed verification profile is reflected in reviewers' relative time allocation. Respondents ranked eight review activities from where they spend the most time (rank 1) to the least time (rank 8) when reviewing merge-candidate HPRs and AIPRs. Because ranks are relative, a lower rank for AIPRs indicates that an activity moved earlier in reviewers' time priorities, necessarily at the expense of other activities.

**Table 4** Mean ranks of review activities for merge-candidate PRs ($N = 239$; 1 = most time, 8 = least). $\Delta$ is AIPR minus HPR; negative values indicate higher priority for AIPRs. All differences except non-trivial logic remain significant after Holm correction.

| Activity | HPR | AIPR | $\Delta$ | $p$ | $\lvert r_{rb} \rvert$ |
|---|---|---|---|---|---|
| Understanding intent from PR description | 2.98 | 4.01 | +1.03 | $< 10^{-14}$ | 0.62 |
| Reasoning about non-trivial logic | 3.18 | 3.18 | 0.00 | .97 | 0.00 |
| Checking tests for risky behavior | 3.87 | 3.38 | −0.49 | $\approx 10^{-7}$ | 0.44 |
| Inspecting surrounding context | 3.86 | 3.42 | −0.44 | $\approx 10^{-8}$ | 0.43 |
| Design and maintainability trade-offs | 4.75 | 5.30 | +0.55 | $\approx 10^{-8}$ | 0.45 |
| Running code locally or exploratory validation | 5.98 | 4.91 | −1.08 | $< 10^{-16}$ | 0.65 |
| Writing review feedback | 5.25 | 6.06 | +0.81 | $\approx 10^{-13}$ | 0.58 |
| Coordinating with others | 6.12 | 5.74 | −0.38 | $\approx 10^{-6}$ | 0.45 |

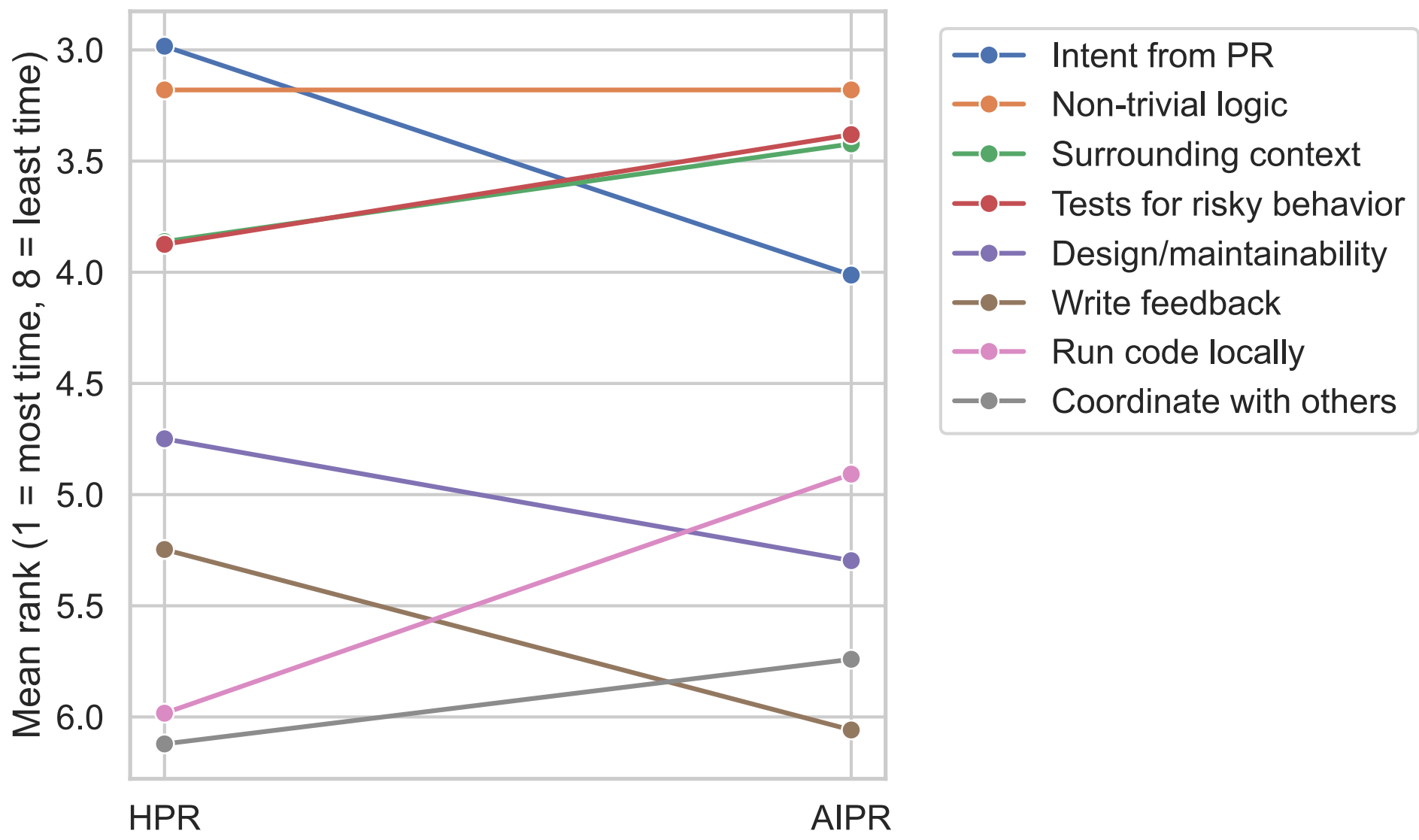

**Fig. 7** Mean rank of review activities for merge-candidate HPRs and AIPRs (1 = most time).

Figure 7 and Table 4 show that AI authorship substantially reordered reviewers' relative time allocation. Compared with HPRs, explanation and communication moved later: understanding intent from the PR description shifted from 2.98 to 4.01 ($\Delta = +1.03$, Wilcoxon $p < 10^{-14}$, $r_{rb} \approx 0.62$), and writing review feedback shifted from 5.25 to 6.06 ($\Delta = +0.81$, $p < 10^{-13}$, $r_{rb} \approx 0.58$). Evidence-producing activities moved earlier. Running code locally or performing exploratory validation shifted from 5.98 to 4.91 ($\Delta = -1.08$, $p < 10^{-16}$, $r_{rb} \approx 0.65$). Checking tests for risky behavior and inspecting surrounding context were also prioritized earlier ($\Delta = -0.49$ and $\Delta = -0.44$, both $p < 10^{-7}$). Coordination with others, including the human prompter for AIPRs, moved slightly earlier ($\Delta = -0.38$, $p \approx 10^{-6}$). Reasoning about non-trivial logic remained central in both conditions, with no meaningful change in mean rank. The construct-level summary reinforces this pattern. A comprehension index, combining intent understanding and feedback writing, moved later for AIPRs (4.12 to 5.04, $\Delta = +0.92$, $p \approx 10^{-21}$, $r_{rb} \approx 0.69$). A defensive-validation index, combining logic reasoning, test checking, context inspection, and local execution, moved earlier (4.22 to 3.72, $\Delta = -0.50$, $p \approx 10^{-19}$, $r_{rb} \approx 0.65$). Governance activities changed only marginally (5.44 to 5.52, $\Delta = +0.08$, $p \approx .03$), suggesting that AI authorship mainly changes how reviewers establish confidence in the submitted artifact rather than uniformly increasing all forms of review effort.

**Finding 3.2:** Even for merge-candidate PRs, respondents reordered relative review effort under AI authorship. Intent comprehension, feedback, and design assessment moved later, while defensive validation, context checking, local execution, and coordination moved earlier.

### 4.4.3 Stop-review conditions

We next characterized when reviewers stop reviewing and require more information before proceeding. Respondents selected the two situations that would most often make them stop reviewing an HPR or AIPR and require additional input.

Figure 8 shows that the dominant stop-review triggers were largely shared across HPRs and AIPRs. Missing or weak tests, unclear intent, non-trivial logic that reviewers could not confidently reason about, and unexplained cross-module or architectural impact were common in both conditions. None of these technical differences remained significant after Holm correction. For example, missing or weak tests was selected by 47.7% of respondents for HPRs and 46.9% for AIPRs, while non-trivial logic remained similarly frequent across conditions (44.8% vs. 46.4%). The clearest authorship-specific shift concerned accountability. Ownership uncertainty doubled as a stop-review trigger, increasing from 14.6% for HPRs to 29.3% for AIPRs (+14.6 pp, Holm-adjusted McNemar $p < .001$). In contrast, the share of respondents who would usually continue reviewing decreased from 13.0% for HPRs to 8.4% for AIPRs, although this difference did not remain significant after correction ($p = .12$ and $p = .19$, respectively).

Open-ended responses further explain this pattern: HPR responses were sparse (14, 5.9%), while AIPR responses were more than twice as frequent (31, 13.0%), suggesting that AI authorship prompted more elaboration of stopping conditions. We manually coded them ($\kappa = 0.97$) into five themes (details in the supplementary material): additional empirical evidence, domain or stakeholder validation, scope and decomposition, provenance and accountability, and review bandwidth or rejection. HPR responses mostly described conventional stopping points, such as bugs, major design issues, missing product approval, lack of accessibility checks, or missing before-and-after evidence. AIPR responses added AI-specific gates, including offline evaluation summaries, sample prediction comparisons, explain plans, benchmark evidence, staging evidence, confirmation that a human checked generated wording, separation of model output from generated code, or splitting broad generated refactors into smaller reviewable pieces. Thus, AIPRs did not broadly change the classic technical reasons for pausing review, but added a stronger accountability layer: reviewers required a human owner, clearer provenance, and stronger evidence before investing further effort.

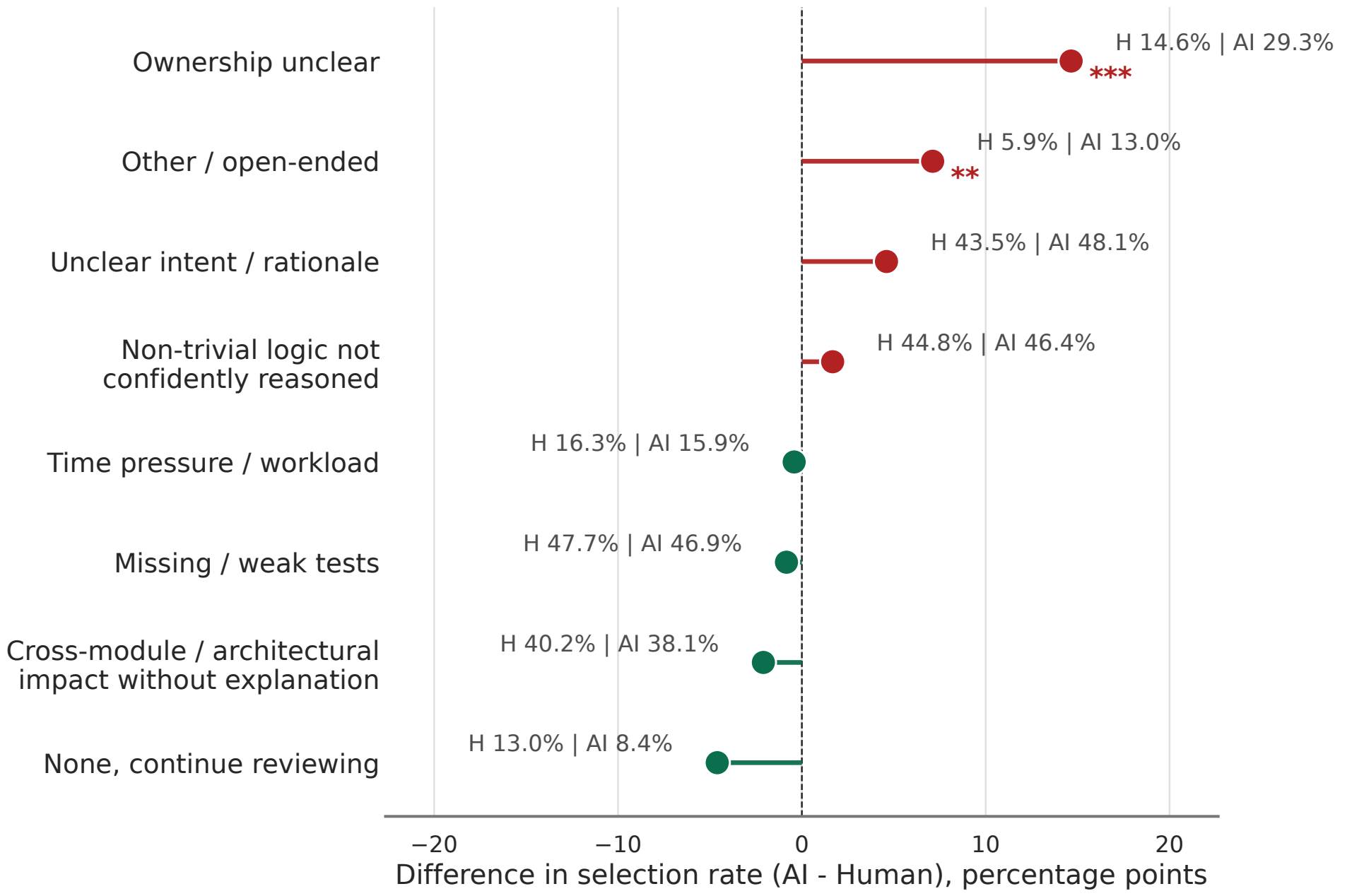

**Fig. 8** Stop-review condition selection rates by authorship condition. Asterisks mark differences that remain significant after Holm correction (exact McNemar tests): ** $p < .01$, *** $p < .001$.

**Finding 3.3:** The main technical triggers for stopping review are similar across HPRs and AIPRs, but AI authorship substantially increases ownership uncertainty and accountability concerns as reasons to pause review.

### 4.4.4 Up-front review-efficiency artifacts

Finally, we examined which up-front artifacts reviewers considered most useful for efficient code review. Respondents selected the two artifacts they would prioritize for HPRs and AIPRs.

Figure 9 shows a pronounced authorship-dependent shift. For HPRs, the dominant artifact was a clear statement of intent and constraints, selected by 70.3% of respondents. It remained the most common AIPR selection, but dropped to 47.7% (−22.6 pp, Holm-adjusted $p < .001$). Minimal reproduction or test-first evidence also decreased from 45.2% to 37.7%, and smaller staged PRs from 40.2% to 33.5%, although neither difference remained significant after Holm correction. In contrast, artifacts that make AI involvement and human accountability explicit became substantially more important for AIPRs. A summary of what was AI-generated versus manually edited increased from 11.7% for HPRs to 35.6% for AIPRs (+23.8 pp, $p < .001$), and explicit ownership statements increased from 16.3% to 33.9% (+17.6 pp, $p < .001$). Other artifacts changed modestly: benchmarks or measurements increased from 18.8% to 21.8%, named maintenance owner remained similar (13.8% vs. 12.6%), and few respondents indicated that none of the artifacts mattered in either condition.

Open-ended artifact responses were provided by 26 respondents for HPRs and 31 for AIPRs. Two raters coded them ($\kappa = 0.94$) into four themes (details in the supplementary material): concrete before-and-after evidence, empirical or domain-specific validation, stakeholder or expert approval, and provenance or accountability. HPR responses mostly described domain-specific evidence that helps reviewers verify changes efficiently, such as screenshots, screen recordings, hardware traces, accessibility checks, offline metrics, example rows, API-version checks, or product and legal approval. AIPR responses included similar evidence requests, but added stronger provenance and accountability requirements: confirmation that generated examples were actually run, which parts were generated versus manually edited, whether a human checked the output, who accepted responsibility for generated wording, and whether generated changes were validated in staging.

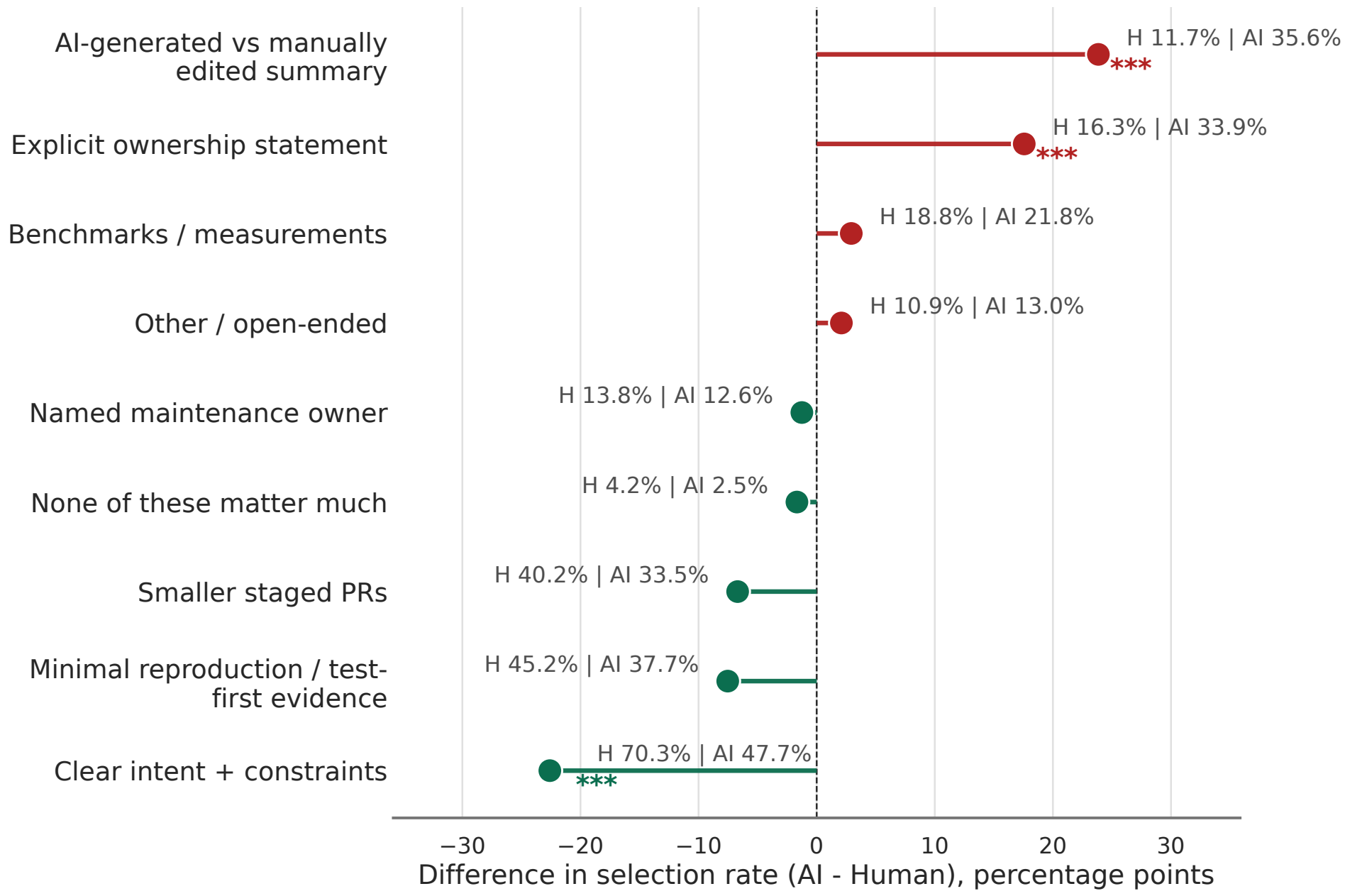

**Fig. 9** Selection-rate differences for review-efficiency artifacts. Asterisks mark differences that remain significant after Holm correction (exact McNemar tests): ** $p < .01$, *** $p < .001$.

**Finding 3.4:** Up-front artifacts shift from general intent and reproducibility for HPRs toward provenance and human accountability for AIPRs. Reviewers are less likely to prioritize clear intent for AIPRs and more likely to require AI-generation disclosure and explicit ownership.

**Answer for RQ$_3$:** Respondents did not describe AIPR review as uniformly broader. Instead, they described a change in the reviewer's role, from interpreting a contribution and communicating with its author to auditing whether a generated change is behaviorally justified, independently demonstrable, and tied to a responsible maintainer. They concentrated effort on inspectable evidence and were more likely to pause when authorship left responsibility ambiguous. The practical need is traceable review support that shows how the change was produced, what has been checked, and who can defend and maintain it.

# 5 Discussion

## 5.1 Evidence Ownership in AIPR Review

Our central interpretation is that reviewing an AIPR raises a question about *who can substantiate the claims behind the change*. A human contributor's history, knowledge of the subsystem, and responsiveness help reviewers decide how much explanation or checking is needed. For an AIPR, those cues do not map directly onto the system that produced the patch or the person who submitted it. Findings 1.2 and 3.1 show where respondents turned instead: to the human prompter rather than to the AI tool, but without extending to the prompter the conversational trust they gave an author, as the drop in clarification requests indicates. The prompter is thus a partial substitute, a person to whom responsibility can be attributed but not necessarily one who can explain the change. This interpretation is consistent with research on appropriate reliance on automation [34, 35] and with Alami et al. [71], who show that code review rests on reciprocal accountability that an LLM reviewer cannot provide. Our study examines the complementary situation in which AI participates as an author, and suggests that the same reciprocity weakens when the change comes from a generator and the accountable person is its prompter.

The distinction between *possessing evidence* and *owning its justification* becomes clearest when RQ$_2$ and RQ$_3$ are read together. Under AI authorship, respondents used tests more as a verification instrument (Finding 3.1) but accepted them less often as the reason a change is sufficiently verified (Finding 2.1). These findings are not inconsistent: tests can be useful for investigating a change without being accepted as sufficient grounds to approve it. An AI-generated test may execute successfully while omitting the behavior that matters, reproducing an incorrect assumption in the patch, or failing to exercise project-specific constraints, concerns that respondents raised in their open-ended answers (Table 3). The survey does not establish that AI-generated tests are generally less reliable. It shows that respondents are less willing to treat test presence as a complete justification under perceived AI authorship.

Evidence ownership is therefore different from requiring an AI-use label. A provenance record can establish how a change was produced, but cannot by itself establish that the behavior was checked against the project's requirements, and a polished PR description may communicate an intended result without showing how it was verified. What respondents sought is evidence that is *traceable to the claimed behavior* and to a person who can answer for it. This is a design principle derived from respondents' stated criteria, not an experimentally tested assurance mechanism. It extends the trust-calibration challenge identified by Gullstrand Heander et al. [72] from deciding where to inspect generated code to deciding whether its supporting evidence warrants approval.

## 5.2 The Hidden Cost of Validation

The findings also point to a redistribution rather than an expansion of review work. Respondents did not select more verification strategies for AIPRs, but they changed which strategies they used and how they ordered their effort (Findings 3.1 and 3.2). The salient cost is therefore not a longer checklist. It is the need to reconstruct an assurance argument that, for a familiar human author, can often rest on established context and conversation.

This suggests a distinction between *validation supplied before review* and *validation reconstructed during review*. When a contributor provides a test result, design rationale, or performance claim without enough information to assess its relevance, the reviewer may need to determine which assumptions were made, whether the test exercises the risky path, and whether the result was obtained in a representative environment. The open-ended responses describing shallow generated tests, missing domain context, and sprawling changes are consistent with this mechanism. Such work can consume reviewer attention even when the diff is readable or the CI pipeline passes. Repository studies of agent-authored PRs observe review interactions and rounds [48, 49]. Our questionnaire adds an account of *which verification activities respondents say they prioritize* and why those activities may be necessary before approval.

The distinction matters when interpreting productivity claims about coding agents. Generating a patch or producing a passing test result does not imply that the associated integration decision becomes cheaper. Conversely, our ranked items do not measure elapsed time, so they cannot show that AIPRs increase total review minutes. The implication is narrower and more actionable: coding agents should be assessed for how much *reviewer reconstruction* their outputs require, alongside generation success [63]. A useful future comparison would measure review time and accuracy with and without independently checkable validation artifacts, separating the cost of reading a diff from the cost of establishing that its supporting claims hold.

The activities that moved later in respondents' priorities are not peripheral. Review feedback is a main channel through which code review transfers knowledge and project norms [9, 10], and design assessment is where reviewers protect long-term maintainability. If validation displaces these activities in practice, AIPR review could check the current change while weakening the knowledge transfer and design scrutiny that review normally provides, a risk echoed in respondents' concerns about unnecessary complexity and duplication in AIPRs (Table 3).

## 5.3 Understanding and Ownership Debt

We use *understanding and ownership debt* as an interpretive concept for a downstream risk, not as a construct directly measured by this survey. It refers to integrating a change without a human who can explain why the implementation is appropriate, identify its governing assumptions, and accept responsibility for subsequent diagnosis or modification. This differs from conventional technical debt, which often concerns compromises in design or implementation [100, 101]. Code can be readable, small, and apparently correct while still lacking an identifiable human who understands the decisions behind it. The debt arises if the knowledge and responsibility needed to maintain that code are not retained after merge.

Three results motivate this interpretation: ownership uncertainty became a more frequent reason to pause review (Finding 3.3), explicit ownership statements and generated-versus-manual summaries gained priority as up-front artifacts (Finding 3.4), and blocking approval became the most common response to unresolved uncertainty (Finding 2.2). Together, these suggest a concern that unresolved uncertainty may persist after integration when no human can explain or defend generated behavior. One further contrast sharpens the interpretation. A *named maintenance owner* was selected at similar rates in both conditions (13.8% for HPRs and 12.6% for AIPRs), whereas the explicit ownership statement, "I can explain and maintain this change," was selected about twice as often for AIPRs. Respondents therefore appear to value evidence of *accepted responsibility for this generated change*, not merely a person's name attached to the PR.

This distinction is particularly consequential when validation requires more than ordinary CI checks. A behavior-preserving refactor requires evidence that externally observable behavior remains stable. Concurrency code demands reasoning about interleavings, security-sensitive changes require assessment of relevant failure modes, and performance claims depend on representative measurement [31, 32, 102–104]. The survey did not track the long-term maintenance outcomes of such changes, so it cannot establish that AIPRs actually accumulate more debt. It identifies conditions under which respondents worry that understanding and accountability will be missing. A longitudinal study linking reviewer evidence, human validation, and post-merge incidents would be needed to test whether this mechanism predicts maintenance outcomes.

## 5.4 Limits of the Authorship Penalty

Lower initial trust should not be mistaken for a universal change in all review criteria. More than one-third of respondents reported equal or higher initial trust in AIPRs (Finding 1.1), about half kept the same sufficient-evidence criterion, verification breadth did not change, and the technical reasons to stop reviewing were similar in both conditions (Finding 3.3). These null and stable patterns are substantive boundaries around the central result: respondents reported pronounced changes in particular judgments, especially sufficient assurance and unresolved uncertainty, without abandoning the technical checks they apply to human-authored code. Although non-significant differences do not establish equivalence, these findings suggest that stricter review practices for AIPRs are not consistent across all dimensions.

This helps reconcile our findings with controlled provenance studies. Huang et al. [69] reported a small difference in acceptance of machine-labeled PRs, Tang et al. [44] found that provenance disclosure affected code-validation behavior, and Singh et al. [70] found no statistically supported effect of a human-versus-AI provenance label on decisions to accept or reject a proposed revision. The studies ask different questions. A label manipulation isolates responses to disclosed source information in a particular task. Our survey asks practitioners to reason about broader HPR and AIPR contexts in which author familiarity, prompt ownership, test credibility, and responsibility may all be considered. The paired comparisons highlight the importance of grounding review decisions in *observable evidence and change risk*, while leaving open the possibility that authorship itself contributes to an unwarranted negative prior. A review process that always demands extra checks from AI-authored contributions, regardless of the evidence already supplied, could waste reviewer effort or disadvantage adequately validated work. The design objective is calibrated reliance rather than categorical distrust [33, 34].

## 5.5 Implications for Review Workflows

The preceding findings suggest a focused design objective: make the basis for approval inspectable *before* reviewers have to reconstruct it. For submitters, this means providing a bounded change and an explanation of the intended behavior in terms of project constraints, together with validation that targets the risky behavior. A human who submits an AIPR should be able to say which parts were generated or revised, which claims were personally checked, what evidence supports those claims, and which questions remain open. A declaration of AI use or ownership alone is not enough. Many projects now adopt AI contribution policies that require disclosure, human involvement, and accountability [54–56]. Our results suggest that such policies should also ask for evidence that lets a reviewer challenge the stated assurance without trusting another generated explanation on its own. At the same time, the ordinary requirements of sound tests, clear intent, and manageable scope remain important for HPRs and AIPRs alike.

For code-review platforms and agentic tools, the most direct opportunity is to present an *evidence trail* connecting each consequential change to its claim, context, validation, and human stewardship. This need not become a universal documentation burden. Routine changes may require only a concise provenance and test summary. Changes involving concurrency, security, data migration, API compatibility, or performance may warrant more explicit assumptions, measurements, or domain-owner review. Tool support could distinguish tests that were merely generated from tests whose expectations were independently checked, identify unverified claims in PR descriptions, and preserve a concise record of human modifications and approvals. These are design propositions consistent with prior work on review information needs and layered review support [11, 39, 72], not tools whose benefits this survey has demonstrated.

For research, the next question is not whether reviewers say they want more evidence, but *which evidence improves their decisions without imposing unnecessary work*. Controlled review studies could extend provenance-label experiments [69, 70] by comparing labels alone with project-grounded evidence packages, assigning conditions between participants or in counterbalanced order so that authorship is not confounded with question order [105], and measuring defect detection, calibrated approval, confidence, review time, and reliance on generated tests. Field studies could examine whether such support changes review interactions and whether accepted AIPRs remain understandable to maintainers over time. Future evaluations should also separate evidence produced by a coding agent from evidence independently validated by a responsible human or another sufficiently independent source. The contribution of this survey is to identify why that distinction matters to reviewers' stated approval logic: plausible code and plausible supporting artifacts are different from a defensible case for merging the change.

# 6 Threats to Validity

## 6.1 Construct Validity

The constructs examined in this study, including trust, evidence requirements, and verification effort, were operationalized through study-specific questionnaire items rather than established psychometric scales. As described in Section 3.2.2, the instrument was grounded in prior research and refined through expert review and pilot testing, and its items ask about concrete judgments and actions, such as the criterion that most often leads a reviewer to approve, rather than abstract attitudes. We therefore interpret individual items as indicators of reviewers' stated judgments rather than as validated latent measures [80]. Confidence in the resulting interpretations comes from convergence across items of different formats and the open-ended explanations. Accountable human ownership, for example, appears as a trust cue, a stop-review condition, an up-front artifact, and an open-ended theme, each elicited by a different question.

An important measurement choice concerns what constitutes an AIPR. The study captures *perceived authorship* rather than verified line-level provenance. In contemporary workflows, AI generation, human prompting, manual editing, and human submission can be intertwined, making a strict boundary between AI-assisted and AI-authored code difficult to observe during review. The questionnaire provided a consistent frame by asking respondents to compare pull requests authored by human developers with those generated primarily by AI systems, by naming the condition in every paired item, and by measuring respondents' experience reviewing AIPRs. The open-ended responses consistently discuss generated code, prompting, human validation, and ownership, indicating that respondents applied the intended distinction. Individual respondents may nevertheless have used somewhat different thresholds for treating a mixed-authorship PR as AI-authored, so we interpret AIPR throughout the paper as a PR that respondents perceived to be primarily AI-generated. AIPR exposure was self-reported in ordinal bands and is therefore used to characterize the sample and for exploratory comparisons rather than for precise measurement.

Predefined response alternatives can make listed considerations more salient than alternatives respondents must generate themselves [84, 106]. Every selection item therefore included an *Other* or *It varies* option with free text, and the open-ended question on AIPR approval independently surfaced concerns about project context, intent, ownership, scope, and maintainability that also appear among the structured options (Table 3). This convergence supports the relevance of the predefined alternatives, though not their exhaustiveness, and we interpret selection frequencies as endorsement of the alternatives presented. Where an option referred to the person responsible for the change, its wording was adapted to each condition (Section 3.2.3). We flag these parallel comparisons, and the main conclusions do not rest on them alone: the approval-criterion shifts toward own reasoning and away from tests, and the verification shifts toward tests and reimplementation, concern identically worded options, and the shift in trust cues is also supported by the change-size cue and by the condition-specific cues.

Some wording intentionally made particular review considerations salient, for example the open-ended question on what makes AIPR approval more difficult and the verification option of applying stricter scrutiny to AIPRs. Such wording allows us to examine how reviewers reason about these considerations, but it may also frame their answers. The questionnaire offered counter-positions, including the same answer in both conditions, *about the same* evidence requirement, *no conscious calibration*, and *usually continue reviewing*, and respondents used them: 27.6% reported about the same evidence requirement for AIPRs, and 29.8% of open-ended responses did not describe AIPR approval as harder. We therefore report the stricter-scrutiny option descriptively and interpret the open-ended themes as reasons given by respondents who described AIPR approval as harder, not as prevalence estimates.

Finally, the study captures respondents' stated judgments rather than behavior reconstructed from repository histories. Self-reports can differ from actual practice and can be affected by recall and social desirability [107, 108]. We mitigated this concern by asking about concrete review situations and actions, and ensuring anonymity. Responses were also not uniformly strict toward AIPRs, as social desirability alone would predict: more than one-third of respondents reported equal or higher initial trust in AIPRs, and about half kept the same approval criterion in both conditions. The findings should therefore be interpreted as evidence about reviewers' stated judgments rather than direct observations of review behavior. Similarly, *understanding and ownership debt* is an interpretive concept synthesized from the responses rather than a directly measured quantity.

## 6.2 Internal Validity

The study is cross-sectional and observational and was designed to characterize reviewer judgments rather than to establish a causal effect of AI authorship on review behavior. The HPRs and AIPRs respondents had in mind may differ in more than authorship, for example in change risk, code quality, contributor familiarity, repository maturity, and organizational AI policy. We therefore avoid interpreting the reported differences as effects caused solely by AI authorship.

The matched HPR and AIPR items strengthen the comparison by asking the same respondent about both conditions with identical stems and response options, which holds constant individual differences in review experience and disposition. However, the two versions of each item appeared on the same page with the HPR version first, and the order was not randomized. The comparison may therefore be affected by order, contrast, or increased authorship salience [109, 110]. We therefore interpret the paired results as within-respondent contrasts under an explicit authorship comparison rather than as causal estimates of authorship in natural review settings.

Because the questionnaire followed a fixed order, earlier items may also have made certain considerations salient when respondents answered later ones. However, the open-ended responses introduced a broader range of concerns than those prompted by preceding items. We cannot rule out context effects, although the observed patterns were not limited to considerations made salient earlier in the questionnaire.

The study description and block introductions made the comparison between human- and AI-authored review explicit. Such transparency was necessary for informed participation, but it may create demand characteristics if respondents inferred the differences the researchers expected [111]. The response patterns were nevertheless not uniformly unfavorable toward AIPRs: a substantial share of respondents answered identically across conditions, selected authorship-neutral options, or reported higher trust in AIPRs, and options concerning conventional technical review, such as missing tests as a reason to stop reviewing, did not differ between conditions (Section 5.4). These patterns provide some assurance against a simple uniform demand effect, but they cannot eliminate it. Our conclusions therefore emphasize the conditional structure of reviewers' judgments rather than treating every HPR–AIPR difference as an isolated effect of authorship. Designs that separate authorship from question order are outlined in Section 5.5.

## 6.3 External Validity

We used complementary recruitment sources to reach practitioners with varied review backgrounds and opportunities for exposure to AI-assisted development: repositories with known agent-authored PR activity, mature Apache and GitHub projects selected independently of AIPR activity, and LinkedIn to reach practitioners beyond public repositories. The final sample contains 239 respondents from 31 countries and spans a wide range of software development and code review experience. Most respondents also reported substantial AIPR review experience, with 78.2% having reviewed at least 10 AIPRs and 49.4% more than 40, providing an experience base relevant to the phenomenon under study.

The recruitment strategy is nevertheless purposive and non-probabilistic. Repository-based recruitment favors practitioners visible in mature open-source projects with publicly available contact information, whereas private repositories, smaller projects, regulated organizations, and non-GitHub ecosystems may be less represented. Professional-network dissemination broadens the sampling frame but also introduces self-selection, and practitioners who chose to participate may differ from non-participants in AI exposure, workload, organizational policy, or strength of opinion. Because anonymous responses cannot be assigned to recruitment channels, a single response rate cannot be computed. We instead report the recruitment process and the achieved sample transparently (Section 3.2.5). Following recommendations for sampling in software engineering, we use the sample to support analytical generalization to comparable review settings rather than statistical estimates for all developers [75, 77].

The findings are also temporally situated. Coding agents, provenance mechanisms, contribution policies, and reviewer familiarity with generated code are evolving rapidly, and exact selection rates may change as tools and governance practices mature. Replication across industrial settings, repository ecosystems, organizational policies, and later generations of coding agents will be needed to establish the longer-term stability and transferability of the reported patterns.

## 6.4 Conclusion Validity

The quantitative analyses were matched to the measurement scale and dependency structure of each item. Paired comparisons used non-parametric or exact tests with effect sizes, Holm correction was applied within families of related tests, and we report response distributions and direction-of-change summaries so that interpretation does not depend on statistical significance alone. The main shifts also remain significant under a Bonferroni correction across all 51 option-level comparisons (largest adjusted $p = .039$). Non-significant findings are interpreted as inconclusive rather than as evidence that HPRs and AIPRs are equivalent on those dimensions.

We also considered response completeness and data quality. From 261 received records, one non-consenting response and 21 responses with more than 50% of the questionnaire unanswered were excluded, resulting in 239 responses for analysis. Optional and open-ended questions were analyzed with item-specific denominators, and missing responses were not interpreted as absence of the corresponding view. Because the survey was anonymous and did not include dedicated attention-check or automated-response detection items, undetected inattentive or machine-assisted responses cannot be ruled out completely. However, participation was not incentivized, most respondents (225 of 239) answered the open-ended question substantively, and all open-ended responses were examined during qualitative coding; responses that were too brief or not interpretable were not assigned substantive themes.

For the qualitative analysis, two coders independently applied a shared codebook with multi-label coding permitted. Disagreements were resolved through discussion, and unresolved cases were adjudicated by a third evaluator. Cohen's $\kappa$ ranged from 0.91 to 0.97, indicating strong agreement in category assignment, while recognizing that qualitative coding remains interpretive. Open-ended survey responses also offer less opportunity for probing and clarification than interviews [112]. We therefore use the qualitative findings to explain, contextualize, and qualify the quantitative patterns, including counterexamples, rather than to claim theoretical saturation or estimate population prevalence.

# 7 Conclusion

AI-authored pull requests challenge a core assumption of modern code review that submitted changes are supported by identifiable humans who can understand, explain, and maintain them. Our survey of 239 practitioners compares reported judgments about human- and AI-authored pull requests. Most respondents reported lower initial trust in AIPRs and placed greater emphasis on the human prompter's expertise and change size than on conventional authorship cues. Before approval, they placed greater importance on independent reasoning than on tests or explanations, were more likely to withhold approval when uncertainty remained, and valued clear ownership statements and code provenance. However, these differences were selective, as technical reasons for stopping review and the range of verification activities remained largely similar. Overall, our findings suggest that AIPR review raises questions about evidence ownership, as reviewers seek credible validation and clear human accountability for generated changes. These findings reflect practitioners' reported judgments rather than directly observed review behavior.

Future research should examine these findings in actual review settings. Controlled experiments could investigate how provenance summaries, ownership statements, and human-validated tests influence approval decisions. Repository and longitudinal studies could further examine their relationships with review practices and post-merge outcomes. As coding agents become more autonomous, research should also investigate how review workflows establish accountability when human involvement in generating changes is limited.

# Declarations

## Funding

This research is supported in part by the Natural Sciences and Engineering Research Council of Canada (NSERC) Discovery Grants Program, the Canada Foundation for Innovation's John R. Evans Leaders Fund (CFI-JELF), and the industry-stream NSERC CREATE program in Software Analytics Research (SOAR).

## Ethical approval

This study received ethics approval from the University of Saskatchewan Behavioural Research Ethics Board (Approval No. BEH#6412) prior to data collection.

## Informed consent

Informed consent was obtained from all participants before participation in the study.

## Author Contributions

Md Shamimur Rahman led the conceptualization and design of the study, developed the survey instrument, conducted participant recruitment and data collection, performed the quantitative and qualitative analyses, interpreted the findings, and drafted the manuscript. Khairul Alam contributed to the qualitative coding and validation of the open-ended responses and reviewed the manuscript. Banani Roy provided research guidance, contributed to the interpretation of the findings, and critically reviewed the manuscript. Chanchal K. Roy supervised the overall research, provided methodological and strategic guidance, contributed to the interpretation of the findings, and critically reviewed and revised the manuscript. All authors reviewed and approved the final version of the manuscript.

## Data Availability Statement

Our replication package and supplementary materials can be found in the following repository: https://doi.org/10.5281/zenodo.20980561

## Conflict of Interest

We have no conflict of interest.

## Clinical Trial Number on the title page

"Not applicable"